\documentclass[aps,pra,showpacs,twocolumn,%longbibliography,
superscriptaddress]{revtex4-1}
\usepackage{amssymb,latexsym,mathrsfs}
\usepackage{amsfonts}
\usepackage{graphicx}
\usepackage{lmodern}
\usepackage{fix-cm}
\usepackage{xfrac}
\usepackage{color}    
\usepackage{textcomp}
\usepackage{amsmath}
\usepackage{subfigure}
\usepackage{float}
\usepackage{datetime}
\usepackage{cancel}
\usepackage{hyperref}
\hypersetup{
    linkcolor=red,
    filecolor=magenta,      
    urlcolor=cyan,
}
\usepackage{tabularx,ragged2e,booktabs}

\newcommand{\bearr}{\begin{array}}
\newcommand{\enarr}{\end{array}}

\newcommand{\bea}{\begin{eqnarray}}
\newcommand{\eea}{\end{eqnarray}}
\def \be{\begin{eqnarray}}
\def \ee{\end{eqnarray}}

\begin{document}

\title{Density profile of noninteracting fermions in a rotating $2d$~trap at finite temperature}

\author{Manas Kulkarni}
\email{manas.kulkarni@icts.res.in}
\affiliation{International Centre for Theoretical Sciences (ICTS-TIFR),
Tata Institute of Fundamental Research,
Bangalore 560089, India}

\author{Pierre Le Doussal}
\email{pierre.ledoussal@phys.ens.fr}
\affiliation{Laboratoire de Physique de l’Ecole Normale Sup\'erieure, CNRS, ENS \& Universit\'e PSL, Sorbonne Universit\'e,
Universit\'e de Paris, 75005 Paris, France}
\author{Satya N. \surname{Majumdar}}
\email{satya.majumdar@universite-paris-saclay.fr}
\affiliation{Universit{\'e} Paris-Saclay, CNRS, LPTMS, 91405, Orsay, France}
\author{Gr\'egory \surname{Schehr}}
\email{schehr@lpthe.jussieu.fr}
\affiliation{Sorbonne Universit{\'e}, Laboratoire de Physique Th{\'e}orique et Hautes Energies, CNRS UMR 7589, 4 Place Jussieu, 75252 Paris Cedex 05, France}

\date{\today}

\begin{abstract}
We study the average density of $N$ spinless noninteracting fermions in a $2d$ harmonic trap rotating with a constant frequency $\Omega$ and in the presence of an additional repulsive central potential $\gamma/r^2$. The average density at zero temperature was recently studied in Phys. Rev. A {\bf 103}, 033321 (2021) and an interesting multi-layered ``wedding cake" structure with a ``hole" at the center was found for the density in the large $N$ limit. In this paper, we study the average density at finite temperature. We demonstrate how this ``wedding-cake'' structure is modified at finite temperature. These large $N$ results warrant going much beyond the standard Local Density Approximation. We also generalize our results to a wide variety of trapping potentials and demonstrate the universality of the associated scaling functions both in the bulk and at the edges of the ``wedding-cake". 
\end{abstract}

\setcounter{page}{1}
\maketitle

\section{Introduction}

Fermions, interacting and noninteracting, in an external confining potential have been a subject of great theoretical and
experimental interest \cite{bloch2008many,nascimbene2010equation,guan2013fermi,cheuk2015quantum,haller2015single,parsons2015site,mukherjee2017homogeneous,hueck2018two}, particularly in the context of cold atoms. In many experimental setups, the interaction between fermions can be tuned and even set to zero, using Feshbach resonances \cite{bloch2008many}. The noninteracting limit is actually far from trivial due to the Pauli exclusion principle. This Pauli exclusion principle often leads to non-trivial spatial distribution of fermions in the presence of an external trap (for reviews see Refs.~\cite{dean2016noninteracting,dean2019noninteracting}). The simplest observable is the macroscopic bulk spatial density that can often be computed using  the well-known local density approximation (LDA)~\cite{butts1997trapped,inguscio2008ultra}
and 
are in principle experimentally observable given the progress in absorption imaging ~\cite{inguscio2008ultra, giorgini2008theory, joseph2011observation}  and quantum gas microscopes~\cite{qm1,qm2,qm3}.
However, the trap leads to sharp edges in the Fermi gas in the limit of a large number of fermions,
and the behavior of the Fermi gas near these edges
are typically not captured by LDA~\cite{dean2016noninteracting,dean2019noninteracting}. 
In one dimension and in some cases in two dimensions,
a large number of recent studies exploited a connection
between the noninteracting trapped fermions at zero temperature and the random matrix theory~\cite{dean2016noninteracting,dean2019noninteracting}. This connection has enabled deriving several
analytical properties of the noninteracting Fermi gas near its edges, going beyond the LDA. These results
extend, beyond the spatial density, to several other
observables, both in one and higher dimensions~\cite{torquato2008point,calabrese2011entanglement,vicari2012entanglement,calabrese2012entanglement, eisler2013universality,marino2014phase,dean2015finite,dean2015universal,calabrese2015random, dean2016noninteracting,dean2019noninteracting,marino2016number,le2016exact,dean2017statistics,le2017periodic,grela2017kinetic,lacroix2017statistics,grabsch2018fluctuations,dean2018wigner,le2018multicritical,schawe2018ground,lacroix2019rotating,stephan2019free,smith2020noninteracting,PhysRevB.101.235169,kulkarni2021multilayered,smith2021counting, smith2022counting}. 

Experiments are always performed at finite temperature. On the theoretical side, however, the connection to RMT holds only at zero temperature. Nevertheless, progress has been made at finite temperature by exploiting the determinantal structure of noninteracting fermions in a trap \cite{dean2015finite,dean2016noninteracting, le2016exact, grela2017kinetic,dean2017statistics,dean2018wigner,grabsch2018fluctuations,dean2019noninteracting,cunden2019free,smith2020noninteracting,smith2021counting}. In particular, it was found that the thermal fluctuations are very relevant, in particular near the edges where the number of fermions are typically very small.  

\begin{figure}[t]
    \includegraphics[width=1.0\linewidth]{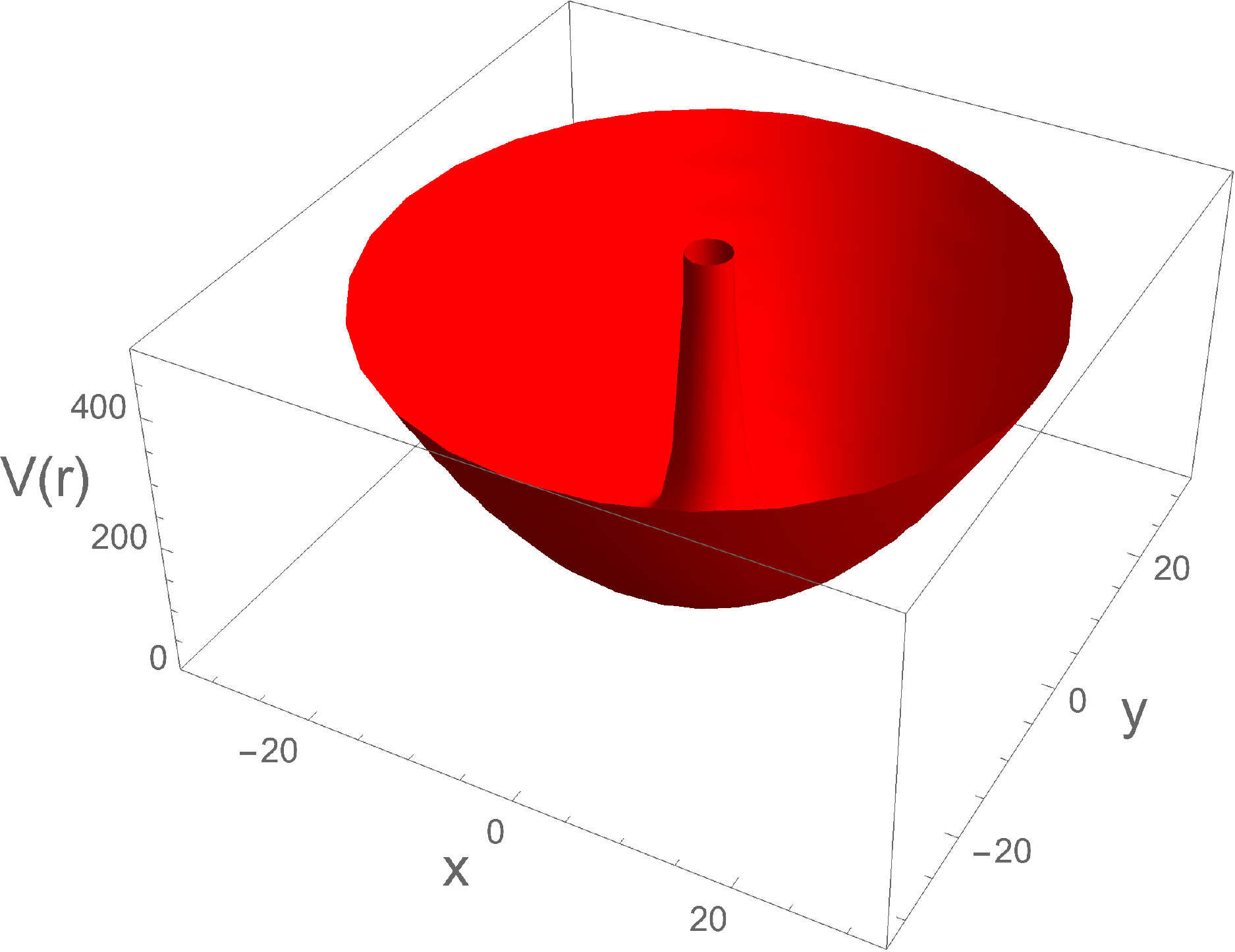} %\quad\quad
  \includegraphics[width=1.0\linewidth]{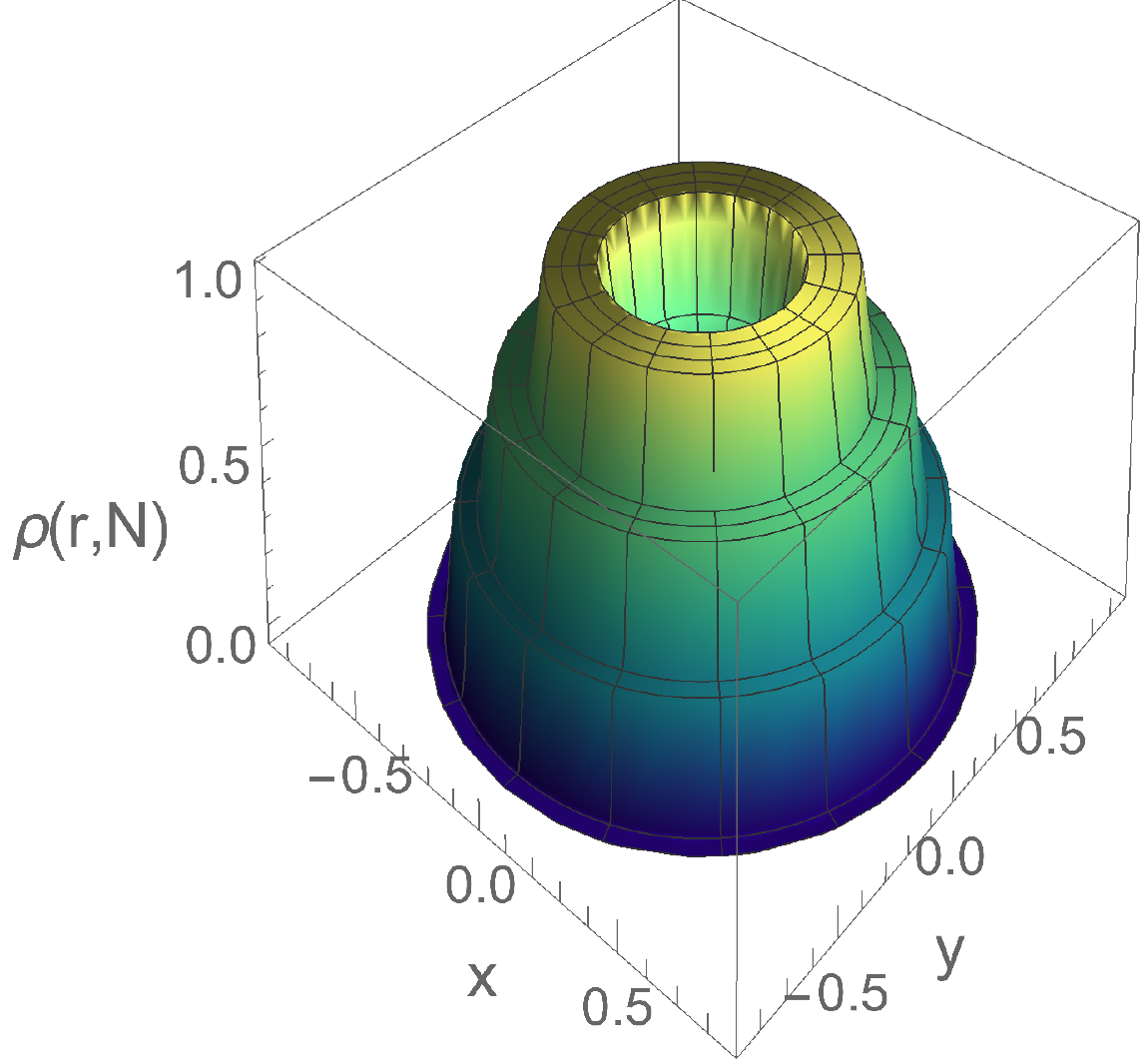}
    \caption{(Top) The external potential $V(r) =  \frac{1}{2} \omega^2 r^2 + \frac{\gamma}{2r^2}$ from Eq.~\ref{def_VM} is plotted in the $2d$ plane. We choose $c=10$ and $N=400$ which means $\gamma = c\,N = 4\times 10^3$. We see the highly repulsive central potential that creates a hole (depleted region devoid of particles) at the center in the density profile shown in the bottom figure. (Bottom)
    A 3D representation of the exact density in Eq.~(\ref{rho_zeroT}). A hole around the origin is surrounded by a multi-layered ``wedding cake'' structure. We choose, $c = 0.5$, $M = 30$ and $N = 8000$ which leads to three layers in this case. The normalization condition Eq.~\eqref{norm_rhoN} gives $\mu = 9.48$.}
    \label{fig:origrhosumschem_S}
\end{figure}

In this paper, we consider a problem of noninteracting fermions in a rotating trap, where the LDA is unable to capture even the bulk density, in addition to the edges \cite{kulkarni2021multilayered,smith2022counting}. Fermions and Bosons in rotating traps have been studied both theoretically~\cite{ho2000rapidly,ho2001bose,aftalion2005vortex,tonini2006formation,cooper2008rapidly, fetter2009rotating, lacroix2019rotating,kulkarni2021multilayered,pereira2022modification} and experimentally~\cite{schweikhard2004rapidly,zwierlein2005vortices}. Here, we consider $N$ noninteracting fermions, with the single particle Hamiltonian in the rotating frame given by \cite{landau1980statistical,leggett2006quantum}
\begin{equation}
\label{ham_introM}
\hat H = \frac{p^2}{2m} + V(r) - \Omega{L}_z\,,
\end{equation}
where $\Omega$ is the frequency of the rotating trap, ${L}_z= x p_y- y p_x = -i  (x \partial_y - y \partial_x)$
is the $z$-component of the angular momentum and $V(r)$ is a confining central potential of the form
\begin{eqnarray} \label{def_VM}
V(r) = \frac{1}{2} m\omega^2 r^2 + \frac{\gamma}{2r^2} \;, \;\; \gamma \geq 0 \;.
\end{eqnarray}
We set the dimensionless ratio $\nu = \Omega/\omega$ in the range $0<\nu<1$, such that the fermions stay confined (when $\nu>1$, the fermions ``fly off" and the system is unstable). In this model, the spatial density of $N$ noninteracting fermions at zero temperature in the limit of large-$N$ was recently studied in Refs.~\cite{kulkarni2021multilayered,smith2022counting} and a very interesting ``wedding cake" structure was unveiled -- this is recalled in Sec.~\ref{sec:finiteN} [see Fig.~\ref{fig:origrhosumschem_S}]. The purpose of this paper is to extend these results to finite temperature. 

We briefly summarize our main results. As in the zero temperature case, we find that for an appropriate large-$N$ limit to exist, one suitable way to
scale the two parameters, $\gamma$ and $\nu = \Omega/\omega$ for large-$N$ is
\begin{eqnarray} \label{def_cM}
c = \frac{\gamma}{N} \; \;\; {\rm and} \;\;\; M = (1-\nu^2)\,N \;,
\end{eqnarray} 
such that $c$ and $M$ are both of order $O(1)$ as $N \to \infty$. We compute the average density of fermions for any finite-$N$ and show how the ``wedding cake" structure seen at zero temperature gets modified. We derive the finite temperature extensions of the bulk and the edge density. We find that the layered structure in the bulk gets affected by increasing temperature only when the temperature is of $O(1)$. In contrast the edge density profile shows visible changes at a much lower temperature of $O(1/\sqrt{N})$. Hence, the edges of the wedding cake are more sensitive to temperature fluctuations than the bulk. We then generalize these results by choosing the confining potential of the form  
\begin{equation}\label{eq:intro_genV}
V(r) = \frac{1}{2}m\omega^2 r^2 + v\bigg(\frac{r}{\sqrt{N}} \bigg)\;,
\end{equation}
where $v(z)$ is a smooth function. We demonstrate that the bulk and edge scaling functions at any finite temperature are universal, i.e., independent of $v(z)$, up to non-universal scale factors that depend on the details of $v(z)$. 

The rest of the paper is organized as follows. In Sec~\ref{sec:finiteN} we recall some zero temperature properties.  In Sec.~\ref{sec:finiteT}, we discuss the density profiles at finite temperature both in the bulk (Sec.~\ref{sec:largeNb}) and the edge (Sec.~\ref{sec:largeNe}). The case of general potential is studied in Sec.~\ref{sec:gen_pot}. We summarise our results along with an outlook in Sec.~\ref{sec:summary}. Certain details are relegated to the appendices.

\section{Recalling the zero temperature properties}
\label{sec:finiteN}
\begin{figure}[t]
\centering
    \includegraphics[width=1.0\linewidth]{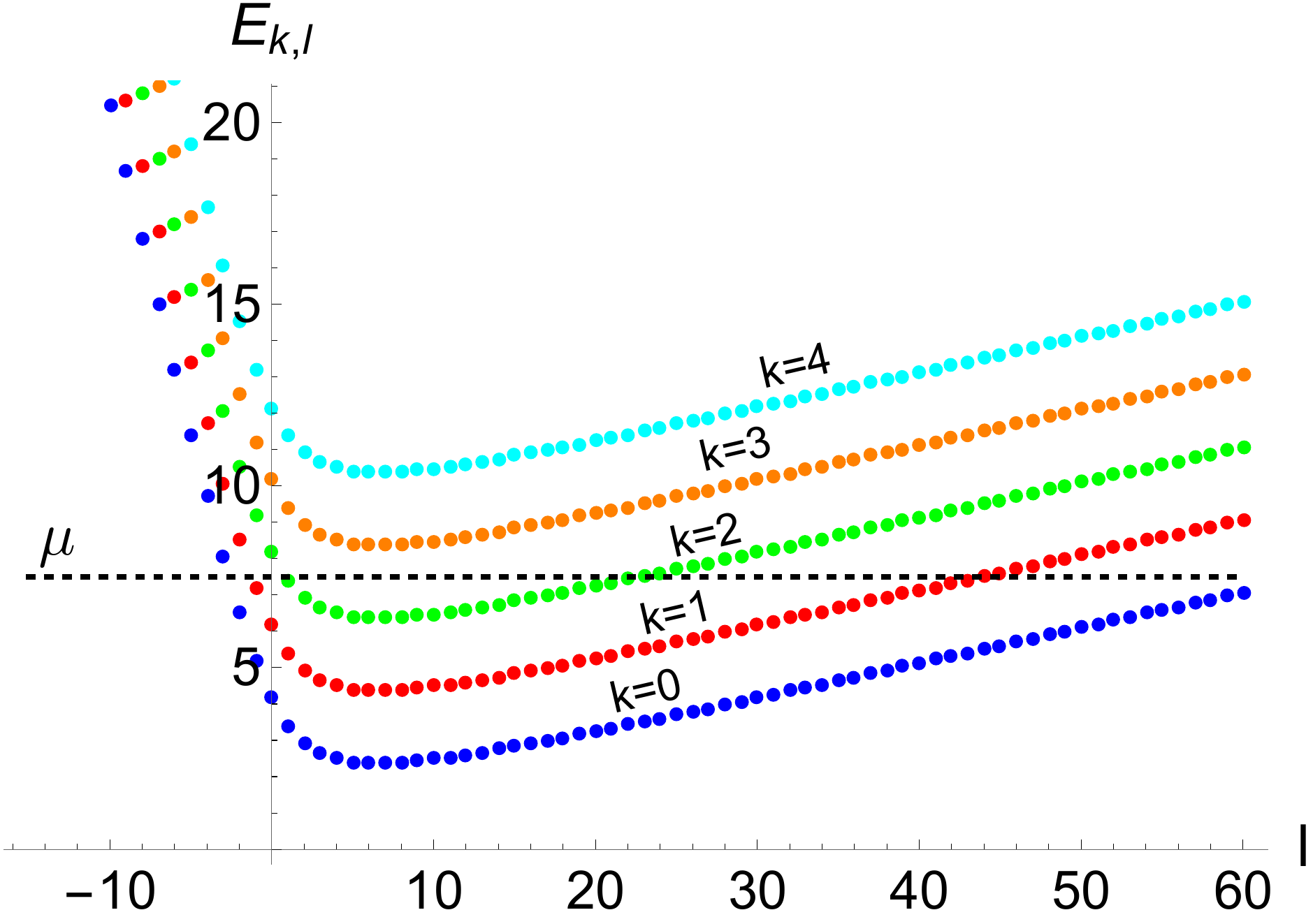}
    \caption{Energy levels $E_{k,l}$ in Eq.~\eqref{eq:meigenvaluesgM} vs $l$ for $k=0, 1, 2, 3,4$, and with the choice of parameters $\gamma = 10$ and $\nu=0.9$. The black (dashed) horizontal line marks the Fermi level, $\mu = 7.5$ (at zero temperature, i.e., when $\beta \to \infty$). Only the states with energy below $\mu$ contribute to the ground state density profile. Hence, in this case, $k^* = 2$ where $k^*$ denotes the highest occupied Landau level. }\label{fig:mdisp}
\end{figure}

In this section, we briefly recall the zero-temperature properties of this model, studied recently in Ref.~\cite{kulkarni2021multilayered}. We start with the Hamiltonian given in Eqs.~\eqref{ham_introM} and \eqref{def_VM}. The eigenfunctions and eigenvalues of this Hamiltonian can be computed exactly in polar coordinates. Setting $m=\hbar=1$, the solutions of the Schr\"odinger equation $\hat H \psi_{k,l}(r,\theta) = E_{k,l} \psi_{k,l}(r,\theta)$ are given by~\cite{kulkarni2021multilayered}  
\begin{equation}
\psi_{k,l}(r,\theta) = a_{k,l} L_k^{\lambda}(r^2) r^\lambda e^{-r^2/2} e^{i l \theta} \;, \; {\rm with} \;\; \lambda = \sqrt{\gamma + l^2} \;,
\label{eq:meigf}
\end{equation}
where $L_{k}^{\lambda}(x)$ are the generalised Laguerre polynomials and the normalisation gives 
\begin{equation}\label{eq:aNorm}
   a_{k,l} = \sqrt{\frac{\Gamma(k+1)}{\pi \Gamma(k+1 + \lambda)}}\;. 
\end{equation}
The associated eigenvalues, in units of $\omega$, are given by 
\begin{equation}
E_{k,l}= 2k+1+\sqrt{\gamma+l^2}- \nu l \;,
\label{eq:meigenvaluesgM}
\end{equation}
where we recall that $\nu = \Omega/\omega$. The single particle states are labelled by a pair of integers $(k,l)$
with $k=0,1,2...$ and $l=0, \pm 1, \pm 2, ...$. The label $k$ is the analogue of the $k$-th Landau level. At zero temperature, due to the Pauli exclusion principle, the fermions occupy the $N$ lowest single particle energy levels of the spectrum in Fig.~\eqref{fig:mdisp}. We denote by $\mu$ the Fermi level, corresponding to the highest occupied single-particle state. For a given $\mu$, we see in Fig.~\ref{fig:mdisp}, that the $k$-th band with $k \leq k^*$ where $k^*$ refers to the highest occupied Landau level, intersects the value $\mu$ at two points $l_{\pm}(k)$. Thus one can write, for all $k \leq k^*$,  
\bea \label{mu_T0}
\mu = E_{k, l_-(k)} = E_{k, l_+(k)} \;.
\eea
At zero temperature, the ground-state many-body wave function $\Psi_0({\bf r}_1, {\bf r}_2, \cdots, {\bf r}_N)$ is given by the $N \times N$ Slater determinant built from the $N$ occupied single particle wave functions  
\bea \label{Slater_T0}
\Psi_0({\bf r}_1, {\bf r}_2, \cdots, {\bf r}_N)  = \frac{1}{\sqrt{N!}} \det_{1 \leq i,j\leq N} \psi_i({\bf r}_j) \;,
\eea
where the index `$i$' refers here to the $i$-th occupied single particle state. The joint probability density of the positions, characterizing the quantum fluctuations at $T=0$, is given by
\bea \label{JPDF_T0}
P_{0}({\bf r}_1, {\bf r}_2, \cdots, {\bf r}_N) = |\Psi_0({\bf r}_1, {\bf r}_2, \cdots, {\bf r}_N)|^2 \;.
\eea
The average density, normalized to $N$, is then given by
\be \label{density_T0}
\rho({\bf r},N) = \sum_{i=1}^N \int \delta({\bf r}-{\bf r}_i) P_{0}({\bf r}_1, {\bf r}_2, \cdots, {\bf r}_N) \prod_{j=1}^N d{\bf r}_j \;, \nonumber \\
\ee 
where the integral in Eq.~\eqref{density_T0} runs over all space. Using the determinantal structure from Eq. (\ref{Slater_T0}), it is easy to show that the average density can be expressed in terms of the occupied single particle eigenfunctions as
\begin{eqnarray} \label{gen_densityMz}
\rho({\bf r}, N) &=& \sum_{k=0}^{k^*} \sum_{l = l_-(k)}^{l_+(k)} |\psi_{k,l}({\bf r},\theta)|^2 \;, %\\ &=& \sum_{k=0}^{\infty} \sum_{l=-\infty}^{\infty} \frac{ |\psi_{k,l}(r,\theta)|^2}{1+ e^{\beta (E_{k,l}-\mu_{\beta})}} \;,
\end{eqnarray}
where the sums over $(k,l)$ run over the $N$ occupied single-particle levels. One can show that this density is isotropic and depends only on the distance $r$ from the center of the trap, i.e., $\rho({\bf r}, N) = \rho(r, N)$. Substituting the explicit expression for the eigenfunctions in Eq. \eqref{eq:meigf} into Eq. (\ref{gen_densityMz}), one gets the exact density profile for any~$N$
\be \label{rho_zeroT}
\rho(r, N) &=& \frac{e^{-r^2}}{\pi} \sum_{k=0}^{k^*} \, \sum_{l = l_{-}(k)}^{l_{+}(k)} \frac{\Gamma(k+1)[L_k^\lambda(r^2)]^2 \; r^{2 \lambda}}{\Gamma(\lambda+k+1) } \;.\nonumber \\
\ee
When plotted as a function of $r$ for fixed $N$, 
this expression in Eq.~\eqref{rho_zeroT} exhibits a ``wedding cake'' structure [see Fig.~\ref{fig:origrhosumschem_S}]. The density has a hole near the center, with radius of order $O(\sqrt{N})$ and the fermions are arranged outside the hole in a layered structure with each 
occupied Landau level contributing to a new layer with decreasing support characterized by $\sqrt{l_{\pm}(k)}$. 
Thus the same $\l_{\pm}(k)$, that characterizes the occupied $k$ levels in the energy space in Eq.~(\ref{mu_T0}) also appear in the expression for the density in the real space~\cite{kulkarni2021multilayered}. For large $N$, using the scaling in Eq.~\eqref{def_cM}, these edges behave as 
\bea \label{l_lambda}
l_{\pm}(k) \approx \lambda_{\pm}(k)\, N  \;,
\eea
where
\begin{eqnarray} \label{lpmk_txt}
\lambda_{\pm}(k) &=& \frac{(\mu - 2 k-1) \pm \sqrt{(\mu-2k-1)^2 - c\; M}}{M} \;,\nonumber \\ 
%k^* &=& {\rm Int}\left[ \frac{\mu - \sqrt{cM}}{2}\right] \;.
\end{eqnarray}
and the Fermi level $\mu \sim O(1)$. Therefore, the density in the $k$-th layer,
expressed in terms of $z=r/\sqrt{N}$ becomes independent of $N$ in the large-$N$ limit and is given by
\begin{eqnarray}
\label{eq:zeroTdensityM}
\rho^{\rm bulk}_k (r=z\sqrt{N}) = \frac{1}{\pi} \mathcal{I}_{\sqrt{{\lambda_-(k)}}}<z<\sqrt{{\lambda_+(k)}}\;,
\end{eqnarray}
where the indicator function $\mathcal{I}$ takes the value $1$ if the inequality in the subscript of Eq.~\eqref{eq:zeroTdensityM} is satisfied and $0$ otherwise. 
The total density is obtained by summing over all $k$ bands upto $k^*$
\begin{equation}
\label{eq:zeroTmain}
  \rho^{\rm bulk} (r=z\sqrt{N}) =  \frac{1}{\pi} \sum_{k=0}^{k^*}  \mathcal{I}_{\sqrt{{\lambda_-(k)}}<z<\sqrt{{\lambda_+(k)}}} \;,
\end{equation}
where we recall that $k^*$ is the highest occupied Landau level. 

In this paper, we ask how this density profile at zero temperature gets modified when a finite temperature is switched on. For example, does the finite temperature destroy this ``wedding-cake" structure ? Surprisingly, we will see that this ``wedding-cake" structure is rather stable with increasing temperature: it essentially gets smeared, as long as the temperature is not too high. However, the density profile near the edges show visible changes at a much lower temperature compared to the bulk. We also find that there is a remarkable universality of the associated scaling functions at finite temperature both in the bulk and at the edges of the ``wedding-cake”. 

\begin{figure}[t]
\centering
    \includegraphics[width=.9\linewidth]{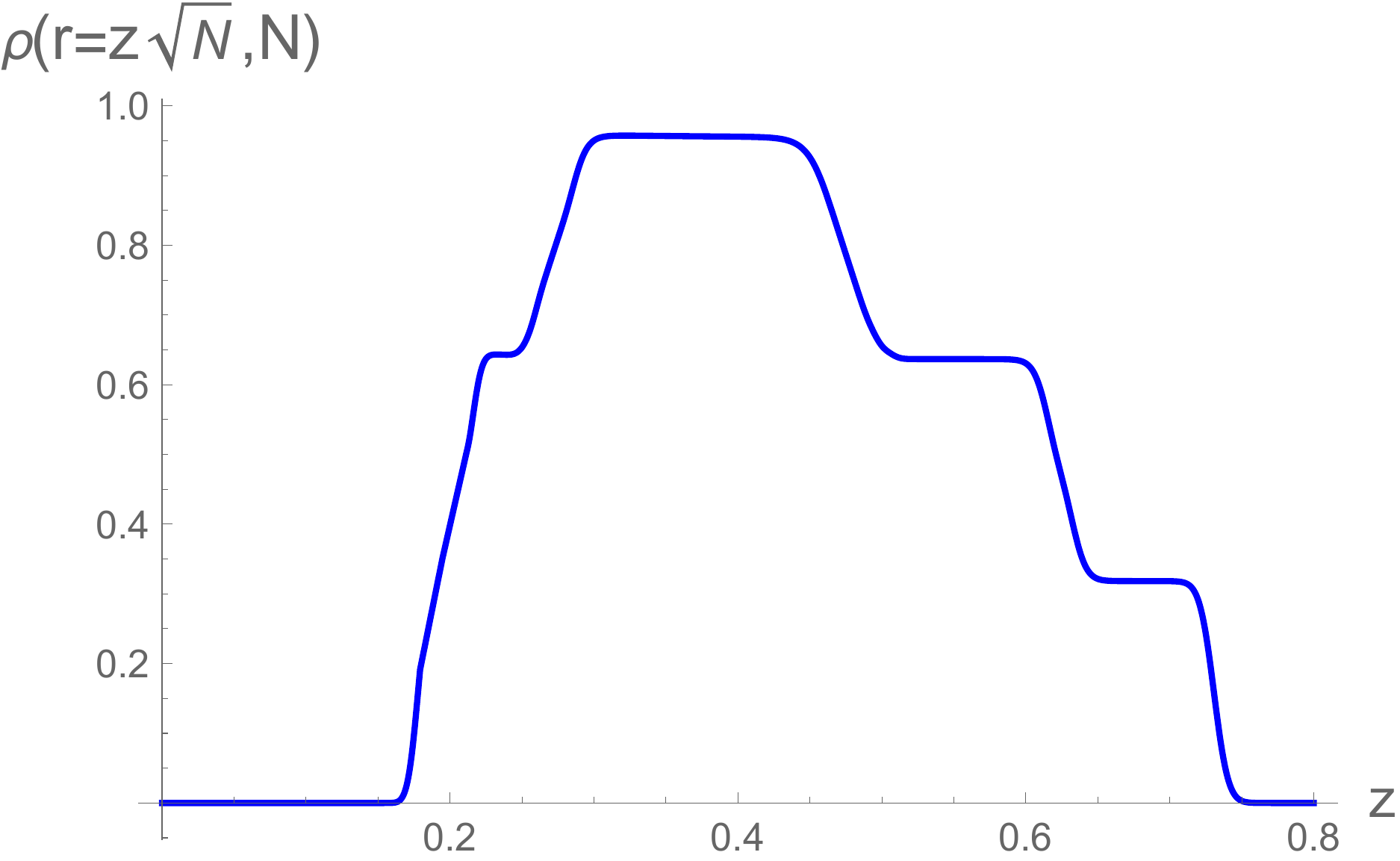}
    \caption{Finite temperature density profile along a radial cut as a function of the scaled radial distance $z = r/\sqrt{N}$ obtained by direct enumeration  of Eqs.~(\ref{gen_densityM}) and \eqref{rho_kM}. The values of parameters used are $c=0.5, \nu = 0.998, \beta=15.0, N=8000$. This is a three layered smeared wedding cake. From the normalization condition [Eq.~(\ref{norm_rhoN})], the corresponding chemical potential turns out to be $\mu_\beta \sim 9.48$. } 
    \label{fig:finiteTnum0}
\end{figure}

\section{Density profile at finite temperature} \label{sec:finiteT}
In this section, we analyse the effects of a finite temperature. For this, let us recall that the joint probability distribution function (PDF) of the positions of the $N$ particles in the canonical ensemble is given by (e.g., see Ref.~\cite{dean2016noninteracting})
\bea \label{P_joint_T}
&&P_T({\bf r}_1, {\bf r}_2, \cdots, {\bf r}_N) = \frac{1}{ Z_N(\beta)} \times \nonumber \\
&&\sum_{E} e^{-\beta E} |\Psi_E({\bf r}_1, {\bf r}_2, \cdots, {\bf r}_N)|^2 \;, 
\eea
where $E$ denotes the energy of a many-body eigenstate $\Psi_E({\bf r}_1, {\bf r}_2, \cdots, {\bf r}_N)$, i.e., a Slater determinant -- similar to Eq.~\eqref{Slater_T0} -- but built from any combination of $N$ single-particle wave functions as given  in Eq.~\eqref{eq:meigf}. One can thus write 
\begin{equation}\label{eq:ME_e}
E = \sum_{k=0}^\infty \sum_{l=-\infty}^\infty n_{k,l} E_{k,l}\;,
\end{equation}
where $E_{k,l}$'s are the single-particle energy levels given in Eq.~\eqref{eq:meigenvaluesgM} and $n_{k,l} = 0,1$ denotes the occupation number of the single-particle state labelled by $(k,l)$. In Eq. (\ref{P_joint_T}),    
the sum over $E$ runs over all possible such $N$-particle states such that 
\begin{equation}\label{eq:norm_nkl}
\sum_{k=0}^\infty \sum_{l=-\infty}^\infty n_{k,l} = N\;, 
\end{equation}
and $Z_N(\beta)$ is the canonical partition function that normalizes the joint PDF $P_T({\bf r}_1, {\bf r}_2, \cdots, {\bf r}_N)$ and it is given by
\bea \label{def_Z}
Z_N(\beta) = \sum_{E} e^{- \beta E} \;.
\eea
In the limit $T \to 0$, i.e., $\beta \to \infty$, the sum in Eq.~(\ref{P_joint_T}) is dominated by the ground-state configuration and it reproduces the $T=0$ result given in Eq. ~\eqref{JPDF_T0}. The average density at finite temperature is then given by Eq.~\eqref{density_T0}, substituting $P_0({\bf r}_1, {\bf r}_2, \cdots, {\bf r}_N)$ by $P_T({\bf r}_1, {\bf r}_2, \cdots, {\bf r}_N)$ given in Eq.~\eqref{P_joint_T}. This is the so-called canonical ensemble, which, however, is hard to analyse for a fixed $N$ due to the hard constraint $\sum_{k,l} n_{k,l} = N$, see, e.g., Ref.~\cite{liechty2020asymptotics}. It is therefore advantageous to work in the grand-canonical ensemble where $N$ is allowed to fluctuate with an additional weight factor $e^{\mu_\beta N}$ where $\mu_\beta$ is the chemical potential. One then determines the chemical potential $\mu_\beta$ from the condition that the total number of particles, {\it on an average}, is given by $N$. In the limit of large $N$, the canonical and grand-canonical ensembles are expected to become equivalent (at least for averaged quantities \cite{grabsch2018fluctuations}). For a detailed discussion, see Ref.~\cite{dean2016noninteracting}. 

Working in the grand-canonical ensemble, the average density can be shown to be given by~\cite{dean2016noninteracting}
\bea \label{rho_GC}
\rho(r,\mu_\beta) = \sum_{k=0}^\infty \sum_{l=-\infty}^\infty  \langle n_{k,l}\rangle |\psi_{k,l}(r,\theta)|^2 \;,
\eea
where
\bea \label{Fermi_fact}
\langle n_{k,l}\rangle = \frac{1}{e^{\beta(E_{k,l}-\mu_\beta)} + 1} \;
\eea
is the Fermi factor. As discussed earlier, one can extract the results for the canonical ensemble from the grand-canonical one by setting the average number of particles to be equal to $N$, i.e.,
\bea \label{rel_N_mub}
N = \sum_{k=0}^\infty \sum_{l=-\infty}^\infty  \langle n_{k,l}\rangle = \sum_{k=0}^\infty \sum_{l=-\infty}^\infty \frac{1}{e^{\beta(E_{k,l}-\mu_\beta)} + 1} \;.\nonumber \\
\eea
Note that, using the isotropy of the average density, the normalization condition [Eq.~\eqref{rel_N_mub}] translates into
\bea \label{norm_rho}
2 \pi \int_0^\infty \rho(r,\mu_\beta) \,r \, dr = N \;.
\eea
Using the relation in Eq.~\eqref{rel_N_mub} connecting $\mu_\beta$ and $N$, the density in the canonical ensemble is then given by 
\bea \label{rho_c_gc}
\rho(r,N) \approx \rho(r,\mu_\beta) \;.
\eea
Consequently, the relation in Eq.~\eqref{norm_rho} implies
\bea \label{norm_rhoN}
2 \pi \int_0^\infty \rho(r, N) \,r \, dr = N \;.
\eea
Injecting the explicit form of the eigenfunctions in Eq.~\eqref{eq:meigf} together with Eq.~(\ref{Fermi_fact}) in Eq.~(\ref{rho_GC}) one finds
\begin{eqnarray} \label{gen_densityM}
\rho(r,N) &=& \sum_{k=0}^{\infty} \rho_k(r, N)  
\end{eqnarray}
where $\rho_k(r, N)$ denotes the contribution to the density from the $k$-th Landau level and is given by
\begin{eqnarray} \label{rho_kM}
\rho_k(r, N) &=& \frac{\Gamma(k+1)\,e^{-r^2}}{\pi} \times \nonumber\\ &&\sum_{l = -\infty}^{\infty} \frac{[L_k^\lambda(r^2)]^2 \; r^{2 \lambda}}{\Gamma(\lambda+k+1) [1+ e^{\beta (E_{k,l}-\mu_{\beta})}]} \;,\nonumber \\
\end{eqnarray}
where we recall that $\lambda = \sqrt{\gamma+l^2}$. 
Note that this result in Eqs.~\eqref{gen_densityM} and \eqref{rho_kM} is exact in the grand-canonical ensemble for any $\mu_\beta$, while it is exact in
the canonical ensemble only in the large $N$ limit -- provided the relation in Eq.~\eqref{rel_N_mub}. In Fig.~\ref{fig:finiteTnum0}, we show the density profile obtained from a direct numerical evaluation of the sums in Eq.~\eqref{gen_densityM} and Eq.~\eqref{rho_kM}. 

In the following, we will analyse this formula [Eq.~(\ref{rho_kM})] first in the ``bulk regime'', where $r \sim O(\sqrt{N})$ and the density is of order $O(1)$. The bulk of this density is supported over a finite region, which we will call the bulk region. At the borders of this bulk regime, the density becomes very small, and we call this region the ``edge regime''. In the two next subsections, we discuss the bulk and then the edge regimes separately.

\begin{figure}[t]
\centering
    \includegraphics[width=1.0\linewidth]{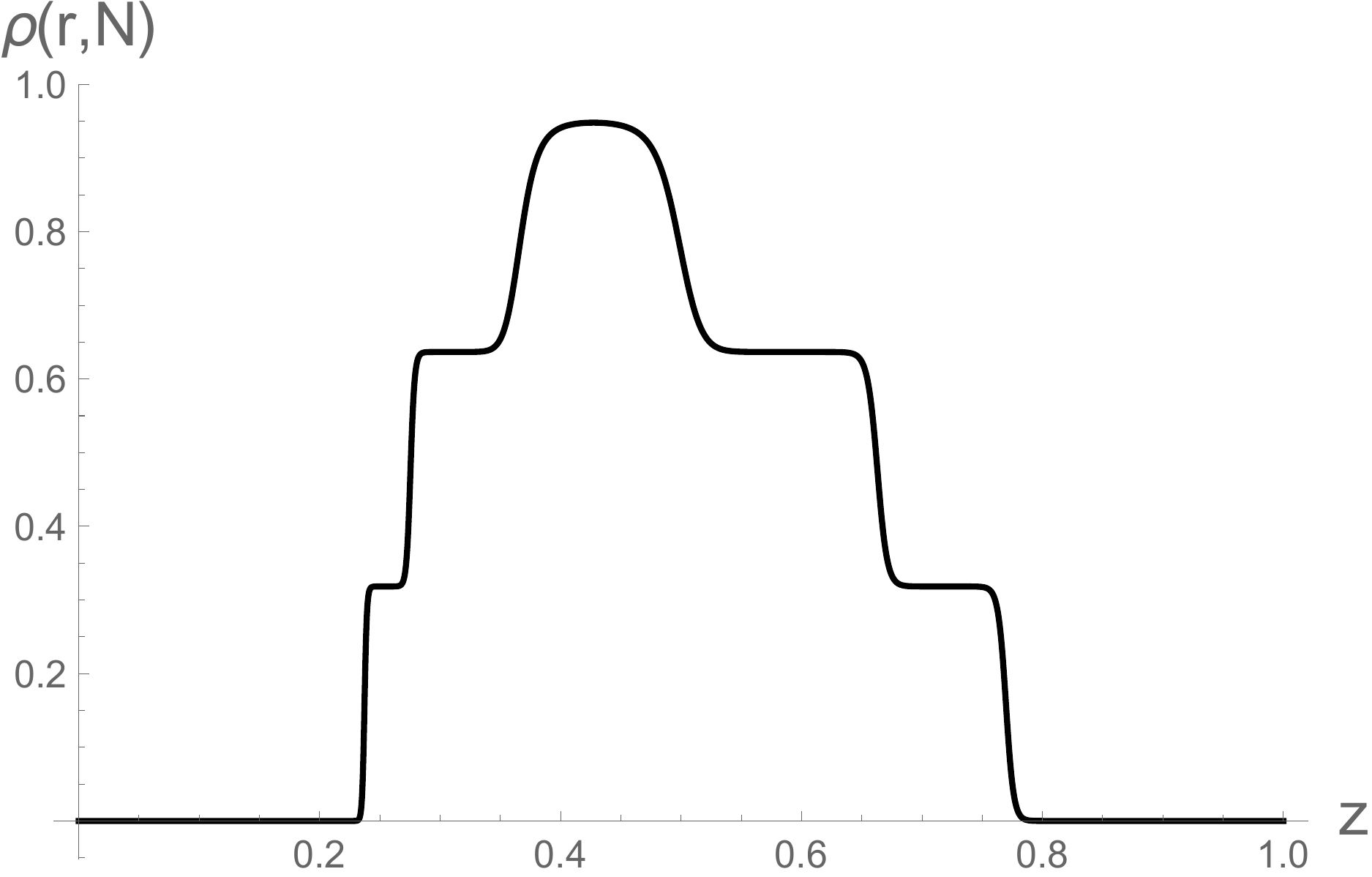}
    \includegraphics[width=1.0\linewidth]{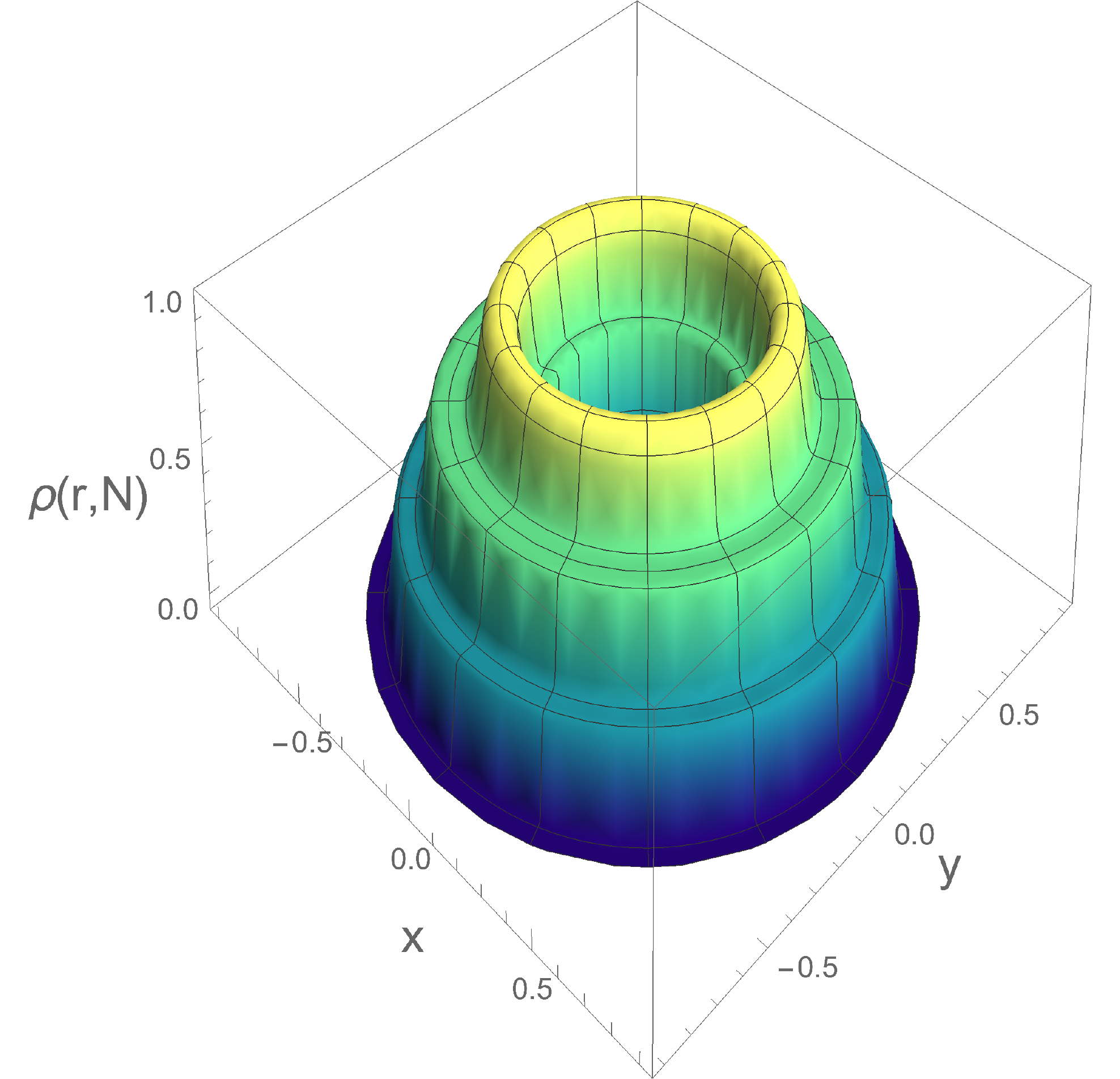}
    \caption{(Top) Finite temperature large-$N$ bulk density profile using Eqs.~(\ref{eq:rhofM}) and (\ref{eq:rhof1M}). The values of the parameters used are $c=1.0$, $M=30$ and $\beta=15.0$.
   This is a three layered smeared ``wedding cake". From the normalization condition [Eq.~(\ref{eq:mu_normM})], the chemical potential turns out to be $\mu_\beta \sim 10.73$.
   (Bottom) A 3D representation of the above large-$N$ bulk density profile. The smearing of the ``wedding cake" can be clearly seen. In the bottom figure, the density profile along a cut in a radial direction is given by the  top panel of the figure. }  \label{fig:finiteTnum}
\end{figure}
\subsection{Bulk regime}
\label{sec:largeNb}
In Appendix~\ref{app:largeNT}, we analyse these Eqs. \eqref{gen_densityM} and \eqref{rho_kM} setting $r = z \sqrt{N}$ and taking the large $N$ limit while keeping $z$ fixed. We show that the bulk density profile takes the scaling form   
\begin{equation}
\rho^{\rm bulk}(r,N) \approx f^{\rm bulk} \bigg(\frac{r}{\sqrt{N}} \bigg),
\label{eq:rhofM}
\end{equation}
where 
\begin{eqnarray}
f^{\rm bulk}(z) &=& \sum_{k=0}^{\infty}f^{\rm bulk}_k(z)  \nonumber \\ 
&=& \frac{1}{\pi} \sum_{k=0}^{\infty}  \frac{1}{1+ e^{\beta (2k+1+\frac{Mz^2}{2}+ \frac{c}{2z^2} -\mu_\beta )}} \;.
\label{eq:rhof1M}
\end{eqnarray}
Note that the normalisation condition in Eq.~\eqref{norm_rhoN} translates into  
\begin{eqnarray}
2\pi \int_{0}^{\infty}  f^{\rm bulk}(z)\,z dz \; =1 \;.
\label{eq:mu_normM}
\end{eqnarray}

\begin{figure}[ht]
\includegraphics[width =\linewidth]{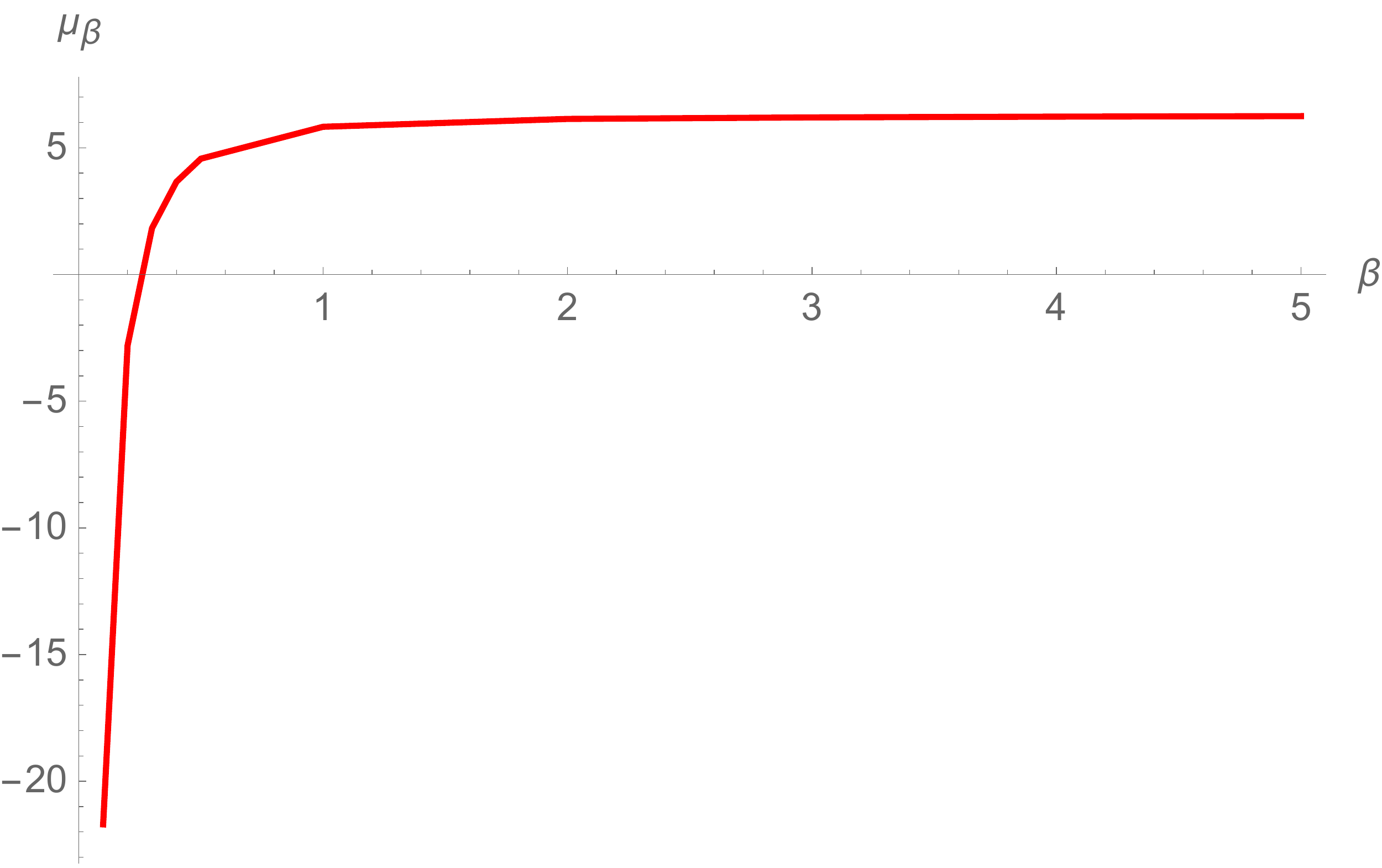}
\caption{Plot of $\mu_\beta$ vs. $\beta$ obtained by solving numerically Eqs.~\eqref{eq:rhof1M} and \eqref{eq:mu_normM}  for $c=1$ and $M = 10$.}
\label{fig:mu_beta}
\end{figure}
Eq.~(\ref{eq:rhofM}) along with Eq.~(\ref{eq:rhof1M}) are part of the main results of this work. From the normalization condition Eq.~(\ref{eq:mu_normM}) and Eq.~(\ref{eq:rhof1M}), it is clear that $\mu_\beta \sim O(1)$ when $\beta \sim O(1)$. The ``Fermi factor" form in Eq.~(\ref{eq:rhof1M}) depends both on the spatial coordinate $z$ and the Landau level $k$ and results in the smearing of the ``wedding cake" that was obtained in Ref.~\cite{kulkarni2021multilayered} at zero temperature. In Fig.~\ref{fig:finiteTnum}, we show a plot of the scaling function $f^{\rm bulk}(z)$ at finite temperature.  

It is worth noting from Eqs.~(\ref{eq:rhofM}) and (\ref{eq:rhof1M}) that, for the bulk density, non-trivial effects due to a finite temperature occur when $\beta \sim O(1)$. This can be understood by the following heuristic argument. In the bulk (recalling that $r = z\sqrt{N}$), in a disk of area $A\sim O(N)$, there are $N_A\sim O(N)$ fermions. This implies that the typical inter-particle spacing between fermions in the bulk is $\bar{a} \sim O(1)$. To understand the relative effects of the quantum versus thermal fluctuations, it is useful to compare this inter-particle spacing to the de-Broglie wavelength associated to a single fermion. If the de-Broglie wavelength is bigger than the inter-particle spacing, quantum effects are dominant, while in the opposite case thermal fluctuations dominate. Since the de-Broglie wavelength scales with temperature as $\lambda_D \propto 1/\sqrt{T}$, we find comparing $\bar{a} \sim \lambda_D$, that the temperature $T\sim O(1)$ in order for the thermal fluctuations to dominate. Thus, beyond this temperature scale, the system starts to  behave classically. 

Let us make a couple of remarks concerning the chemical potential $\mu_\beta$ which is obtained from the normalization condition in  Eq.~\eqref{eq:mu_normM}. For $\beta \to \infty$ (,i.e., $T \to 0$) the chemical potential reaches a constant $\mu_\infty$, which coincides with the Fermi level $\mu$ given in Eq. (\ref{mu_T0}), which translates [using Eq. (\ref{eq:meigenvaluesgM})] into
\bea \label{mu_inf}
2 k +1 +\frac{M}{2}\lambda_{\mp}(k)+\frac{c}{2\lambda_{\mp}(k)} -\mu_{\infty} = 0 \;.
\eea

When we increase the temperature, i.e., reduce $\beta$, the chemical potential $\mu_\beta$ remains extremely robust and stays quite close to $\mu_\infty$ [see Fig.~(\ref{fig:mu_beta})]. This is why, even for a reasonable finite temperature the notion of layers seen in the zero temperature case still persists. 
%When $T \sim O(1)$, then we notice that $\mu_\beta$ starts reducing and finally goes all the way to $-\infty$ as $T\to \infty$. In Appendix~\ref{app:subsecR} we discuss more details and show that   
In Appendix~\ref{app:subsecR}, we derive the asymptotic behavior of $\mu_\beta$  
\begin{eqnarray} \label{mub_asympt}
\mu_\beta \approx 
\begin{cases}
&\mu_{\infty} - \frac{1}{\rm Int \big[ \frac{\mu_\infty}{2} \big]} \frac{1}{\beta}\, e^{-A\, \beta}\;,\quad \text{when} \,\,\beta \to  \infty \\
& \\
& 2\frac{\log(\beta)}{\beta}, \quad \quad \quad \quad \hspace*{1.2cm}\text{when} \,\,\beta \to 0 
\end{cases}
\end{eqnarray}
where $A$ is a constant, i.e., independent of $\beta$, given in Eq. (\ref{A_app}). 

It is easy to see that, in the $T\to 0$ limit, the expression for the density in Eqs.~(\ref{eq:rhofM}) and (\ref{eq:rhof1M}) give back the ground-state result obtained in Ref.~\cite{kulkarni2021multilayered}. Indeed, in this limit the ``Fermi-factor" forms reduce to indicator functions. On the other hand, in the high temperature limit, one recovers the classical Gibbs-Boltzmann distribution for independent particles in the external potential $V(r = z \sqrt{N})$ (see Appendix~\ref{app:largeNT} for details). 
 
We conclude this subsection by discussing the small and large arguments behavior of the scaling function $f^{\rm bulk}(z)$ in Eq.~(\ref{eq:rhof1M}). 
%The small-$z$ behaviour will describe how the density diminishes as one goes to the centre of the smeared wedding-cake. 
In the limit $z \to 0$, it is easy to see that $f^{\rm bulk}(z)$ has an essential singularity, 
%which to leading order is given by the $k=0$ term,
leading to 
\begin{equation}
f^{\rm bulk}(z) \approx B_\beta \, e^{-\frac{\beta c}{2z^2}},\, \text{for } z \to 0 \;,
\label{eq:rhofsmallzM}
\end{equation}
where 
\be \label{Bbeta}
B_\beta = \frac{e^{\beta \mu_\beta}}{2\pi \sinh(\beta)}\;.
\ee
%\blue{recheck and simplify} 
On the other hand, for large $z$, one finds
\begin{equation}
f^{\rm bulk}(z)  \approx B_\beta \, e^{-\frac{\beta M z^2}{2}},\, \text{for } z\to  \infty \;,
\label{eq:rhoflargez}
\end{equation}
with the same amplitude $B_\beta$ given in Eq.~\eqref{Bbeta}. 

%\newline
\subsection{Edge regime}\label{sec:largeNe}
    
We will now investigate the behaviour of the density at the ``edges" of the smeared ``wedding cake". At $T=0$, the position of the left and right edges associated to the $k$-th level are located at $\sqrt{l_{\mp}(k)} \approx \sqrt{\lambda_{\mp}(k) N}$ where
\begin{eqnarray} \label{lpmk_txte}
\lambda_{\pm}(k) &=& \frac{(\mu_\infty - 2 k-1) \pm \sqrt{(\mu_\infty-2k-1)^2 - c M}}{M}\;.\nonumber \\ 
%k^* &=& {\rm Int}\left[ \frac{\mu - \sqrt{cM}}{2}\right] \;.
\end{eqnarray}
We start with the exact expression for $\rho_k(r,N)$ in Eq.~(\ref{rho_kM}) and set 
\begin{equation}
r \approx  \sqrt{\lambda_{\mp}(k) N} + \frac{u}{\sqrt{2}},\quad \text{at the left/right edge} \;,
\label{eq:zuedger01M}
\end{equation}
where $u \sim O(1)$ denotes the distance from the left/right edge at $T=0$. We find that for large-$N$ (see Appendix~\ref{app:edge} for details), the density can be approximated as 
\bea \label{rho_edge_S_FTM}
\rho_k^{\rm left/right   \,\,edge}(r,\theta,N) \to  f^{\rm left/right\,\, edge}_k(u,N)\;,%\;, \; {\rm where} \;\; f_k^{\rm edge}(u) = \frac{2^{-k}}{\pi^{3/2} \Gamma(k+1)} \int_{-u}^{\infty} dx \, e^{-x^2} \, [H_k(x)]^2 \;}
\eea
with 
\begin{widetext}
\begin{eqnarray} \label{eq:fedge_FTM}
f^{\rm left/right\,\,edge}_k(u, N) = \frac{2^{-k}}{\pi^{3/2} \Gamma(k+1) }  \int_{-\infty}^{+\infty} dx \, \frac{e^{-x^2} \, [H_k(x)]^2}{{1+ e^{\beta \big(2 k +1 +\frac{M}{2}\lambda_{\mp}(k)+\frac{c}{2\lambda_{\mp}(k)} -\mu_{\beta}\big)} e^{\big( \frac{\beta}{\sqrt{2N}}(u+x) \sqrt{\lambda_{\mp}(k)} \big[M - \frac{c}{\lambda_{\mp}(k)^2} \big] \big)}}}\;,
\end{eqnarray}
%\label{eq:fedge_FT}
\end{widetext}
where, again, the $\mp$ subscript refers to the left/right edges and where {$H_k(x)$} is the Hermite polynomial of index $k$.

From Eq.~(\ref{eq:fedge_FTM}), we note that in order that the edge density approaches an $N$-independent scaling form for large $N$, one needs to scale $\beta \sim \sqrt{N}$. In this case, the argument of the second exponential in the denominator becomes independent of $N$. Additionally, when $\beta\sim\sqrt{N}$, the argument of the first exponent in the denominator in Eq.~(\ref{eq:fedge_FTM}) vanishes due to Eq.~\eqref{mu_inf} and Eq.~\eqref{mub_asympt}. The $N$-independent edge scaling function is then given by 
%\begin{widetext}
\begin{eqnarray} \label{eq:fedge_FTMsimp}
f^{\rm left/right\,\,edge}_k(u) = \frac{2^{-k}}{\pi^{3/2} \Gamma(k+1) }  \int_{-\infty}^{+\infty} dx \, \frac{e^{-x^2} \, [H_k(x)]^2}{{1+ e^{ b_{\mp}\,(u+x) }}} \,,\nonumber \\
\end{eqnarray}
%\end{widetext}
where
\begin{equation}
\label{eq:bpm}
    b_{\mp} = \tilde{\beta}\, \sqrt{\lambda_{\mp}(k)} \,\left(M - \frac{c}{\lambda_{\mp}(k)^2} \right) \;,
\end{equation}
with $\tilde \beta = \beta/\sqrt{2N}$. Thus $b_{\mp}$ can be interpreted as an ``effective inverse temperature". We show in Appendix~\ref{app:edge} that $b_{-}<0$ and $b_{+}>0$ corresponding respectively to the left and right edges of the $k-$th layer. 
%\label{eq:fedge_FT}

\begin{figure}[t]
\centering
    \includegraphics[width=.9\linewidth]{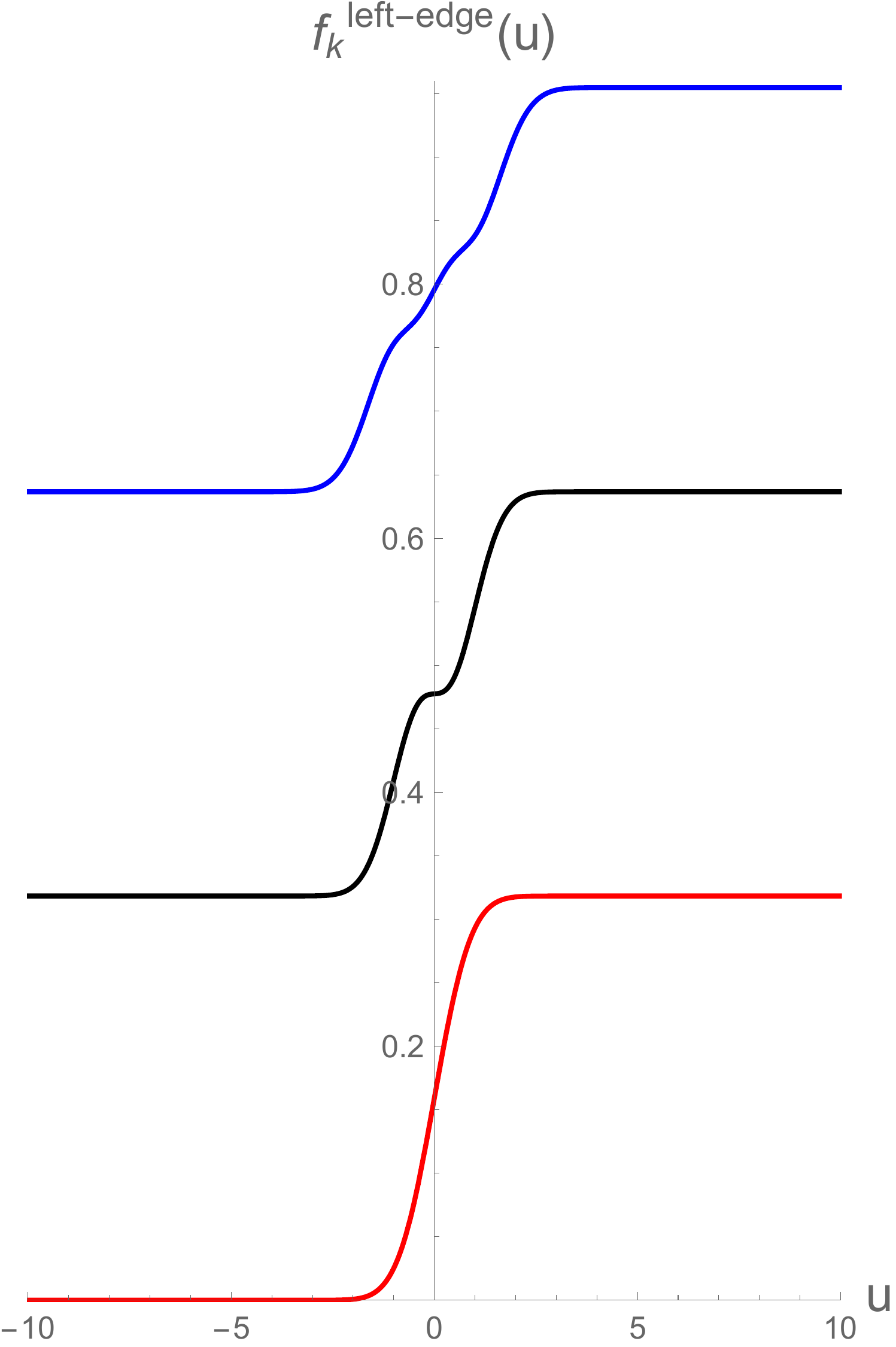}
    \caption{Finite temperature large-$N$ edge density profile (for the left edge) using Eq.~(\ref{eq:fedge_FTMsimp}) for $k=0,1,2$ (red, black and blue respectively).
    The values of parameters used are $c=1.0, M = 30, \beta = 50, N =4000, \tilde{\beta}= \beta/\sqrt{N}=0.56$. 
    From normalization condition [Eq.~(\ref{eq:mu_normM})], the corresponding chemical potential is $\mu_\beta \approx 10.73$. Note that the plots are vertically translated by $1/\pi$ for a better visualization.} \label{fig:finiteTedge}
\end{figure}

 Our results for the edge [Eq.~\eqref{eq:fedge_FTMsimp}] is plotted in Fig.~\ref{fig:finiteTedge}. Also, one can verify that the zero temperature limit of Eq.~(\ref{eq:fedge_FTMsimp}) reproduces the result in Ref.~\cite{kulkarni2021multilayered} and this is detailed in Appendix~\ref{app:edge}.  

We conclude this section by providing a heuristic explanation for the non-trivial finding that $\beta\sim O(\sqrt{N})$ for edge effects to be visible.
%In other words, temperature should be less than $O(1/\sqrt{N})$. This is a non-trivial finding of our calculations and we try to provide a heuristic explanation. 
For this purpose, it is useful to rewrite the Hamiltonian in  Eq.~(\ref{ham_introM}) in the following form~\cite{smith2022counting},
\begin{eqnarray}
\label{ham_intro_pa}
\hat H &=&  \frac{1}{2}(p-\textbf{A})^2+ \frac{\gamma}{2r^2}  + \frac{1}{2} \omega^2 r^2 (1 -\nu^2)\,, \nonumber \\
%&\approx&
\end{eqnarray}
%\appendix
where $\textbf{A}$ is the vector potential $(\Omega y,-\Omega x) $. For our discussion here, we can set
$\gamma = 0$ since it turns out not to have any consequence. %Hence, we set this to zero (i.e., $c\to 0$). 
Note that when there is no rotation (i.e., $\Omega\to 0, \nu \to 0$), the density forms a ``cap"~\cite{dean2016noninteracting}. Without rotation, the inverse temperature scale at the edge turns out to be $\beta \sim N^{1/6}$ (from Ref. \cite{dean2015universal}) which certainly suggests that the scale $\beta \sim N^{1/2}$ owes its origin to the rotation. 

The edge sensitivity turns out to be solely due to rotation. To see this, 
we analyze the last term in Eq.~(\ref{ham_intro_pa}). 
Using Eq.~(\ref{def_cM}) this last term simplifies to 
\begin{eqnarray}
\label{ham_intro_pa_last}
 \frac{1}{2} \omega^2 r^2 (1 -\nu^2) \approx \frac{M}{2N} \omega^2 r^2 \;.
\end{eqnarray}
Furthermore, near the edge we have using Eq.~(\ref{eq:zuedger01M})
\begin{equation}
r^2 = \lambda_+(k) N + \sqrt{2  \lambda_+(k) N} u \;.
\label{eq:zuedger}
\end{equation} 
where $u \sim O(1)$. Hence, Eq.~\eqref{ham_intro_pa_last}, in the edge regime, reduces to 
\begin{eqnarray}
\label{ham_intro_pa_lastsimp}
 \frac{1}{2} \omega^2 r^2 (1 -\nu^2) \approx \frac{M \omega^2}{2}\lambda_+(k)+ \frac{M\omega^2}{\sqrt{2}}\sqrt{\lambda_+(k)}\frac{u}{\sqrt{N}}  \;.\nonumber \\ 
\end{eqnarray}
Substituting Eq.~\eqref{ham_intro_pa_lastsimp} in the Hamiltonian in Eq.~\eqref{ham_intro_pa} one can see that to sustain a spatial fluctuation $u\sim O(1)$ at the edge, the cost of energy $\Delta H \sim 1/ \sqrt{N}$. Comparing it to the thermal energy of order $\beta = 1/T$, one gets the temperature scale  $T \sim  1/ \sqrt{N}$. If $T >  1/\sqrt{N} $, then the thermal fluctuations will be too high and will wash away the effect of finer quantum fluctuations in Eq.~(\ref{eq:fedge_FTM}). Hence, for the zero temperature edges to be still visible, one can raise the temperature only up to $O(1/\sqrt{N})$, and not more than that. 

We end this section by mentioning that a high temperature regime, where $\beta \sim 1/\sqrt{N}$ (hence
quite different from the regime studied here where $\beta \sim \sqrt{N}$) was considered in Ref. \cite{garcia2002critical}. It would be interesting to study the crossover between these two regimes

\section{Generalization to a wide class of potentials} \label{sec:gen_pot}
In this section, we will show that our procedure can be adapted to a wide class of potentials and not merely restricted to external potentials of the form in Eq.~\eqref{def_VM}. This is also experimentally relevant since different trapping potentials can be engineered. We consider a wide class of potentials of the form
\begin{equation}
V(r) = \frac{1}{2}m\omega^2 r^2 + v\bigg(\frac{r}{\sqrt{N}}\bigg)\;, 
\label{eq:vgenM}
\end{equation}
where $v(z)$ can be an arbitrary smooth function. In previous sections, we had focused on a specific case with $v(z) = c/2z^2$. 

It turns out that both the bulk and the edge density profiles can be computed for general potentials of the form in Eq.~(\ref{eq:vgenM}) [for details, see Appendix.~\ref{app:generalP}]. 
The bulk scaling function for the density
[the analogue of Eq.~(\ref{eq:rhof1M})] turns out to be 
\begin{eqnarray}
f^{\rm bulk}(z) &=& \sum_{k=0}^{\infty}f^{\rm bulk}_k(z)  \nonumber \\
&=& \frac{1}{\pi} \sum_{k=0}^{\infty}  \frac{1}{1+ e^{\beta (2k+1+\frac{Mz^2}{2}+ v(z) -\mu_\beta )}} \;.
\label{eq:rhof1Mgen}
\end{eqnarray}
Note that, in the zero temperature limit Eq.~\eqref{eq:rhof1Mgen} becomes
\bea 
\label{eq:maingenT0}
f^{\rm bulk}(z) &=& \frac{1}{\pi}   \sum_{k = 0}^{\infty} 
\, \theta \bigg( \mu - \Big(2 k+1 + \frac{M}{2} z^2 + v(z) \Big) \bigg) \;, \nonumber \\ % \int dx |\psi_k(x)|^2 \nonumber \\ 
\eea 
where $\theta(x)$ is the Heaviside theta function. For the specific case of $v(z) = c/2z^2$ we recover Eq.~\eqref{eq:zeroTmain} using Eq.~\eqref{lpmk_txt}.

Similarly, the edge density can also be generalized. To do so, we need to expand about the corresponding edges $z_e$ of $T=0$ which are given by the solutions to the equation 
\begin{eqnarray}
\label{edgepositionsM} 
\frac{M}{2} z_e^2 + v(z_e)  + 2 k + 1  =  \mu_\infty.
\end{eqnarray}
After some algebra (see Appendix~\ref{app:generalP} for details) it turns out that Eq.~\ref{eq:fedge_FTMsimp} gets generalized to 
\begin{widetext}
\begin{eqnarray} \label{eq:fedge_FTMGen}
f^{\rm left/right\,\,edge}_k(u) = \frac{2^{-k}}{\pi^{3/2} \Gamma(k+1) }  \int_{-\infty}^{+\infty} dx \, \frac{e^{-x^2} \, [H_k(x)]^2}{{1+ e^{ \tilde{\beta}(u+x) \big(Mz_e+v^{\prime}(z_e)\big) }}},
\end{eqnarray}
%\label{eq:fedge_FT}
\end{widetext}
where $\tilde{\beta} = \beta/\sqrt{2N}$. The zero temperature limit for the edge density profile in the case of general potential is detailed in Appendix.~\ref{app:generalP}.

\section{Summary and Outlook}
\label{sec:summary} 
In this paper, we studied $N$ noninteracting fermions in a two dimensional rotating trap in a general class of confining potentials at finite temperature. We focused on the average density of fermions and computed the universal large-$N$ density both in the bulk and edge regimes. We believe that these results will stimulate experimental measurements which can confirm our findings. 

It is interesting to investigate how new layers/droplets at finite temperature get formed (nucleation) when one varies parameters in the problem \cite{in_preparation,kulkarni2021multilayered}. 
It would be interesting to explore the 
finite temperature properties of other observables going beyond the average density.
These include, for instance, the two-point correlation functions and the full counting statistics of a given domain. The full counting statistics have been recently computed exactly at zero temperature for (i) rotating noninteracting fermions~\cite{lacroix2019rotating,smith2022counting} and related Ginibre matrix ensembles  \cite{charlier2021asymptotics,charlier2022exponential} and (ii) for a class of interacting $1d$ fermions exploiting connections to random matrices~\cite{smith2021full}. It will also be interesting to explore how the universal structures found here for noninteracting fermions get affected in the presence of interactions.  It will be extremely interesting to study the non-equilibrium dynamics of this system when subject to quenches. For example, in the temperature quench one can suddenly cool the system (to zero temperature) and ask how the finite temperature density evolves in time. Also, one can quench the trapping frequency $\omega$ (say $\omega \to 2\omega$) of the Harmonic trap in Eq.~\eqref{def_VM} and study the non-equilibrium dynamics. \\

\acknowledgements

We are grateful to N. R. Smith for discussions and ongoing collaboration on related topics.
M. K. would like to acknowledge support from the Project 6004-1 of the Indo-French Centre for the Promotion of Advanced Research (IFCPAR), Ramanujan Fellowship (SB/S2/RJN114/2016), SERB Early Career Research Award (ECR/2018/002085) and SERB Matrics Grant (MTR/2019/001101) from the Science and Engineering Research Board (SERB), Department of Science and Technology, Government of India. M. K. acknowledges support from the Department of Atomic Energy, Government of India, under Project No. RTI4001. M. K. thanks the hospitality of  \'Ecole Normale Sup\'erieure (Paris) and Le Laboratoire de Physique Théorique et Modèles Statistiques (LPTMS, Université Paris-Saclay). M. K. acknowledges support from the Infosys Foundation International Exchange Program at ICTS. This research was supported by ANR grant ANR-17-CE30-0027-01 RaMaTraF. We also acknowledge hospitality and support from  Galileo Galilei Institute,
and from the scientific program on ``Randomness, Integrability, and Universality".
\\

\appendix

\section{Bulk density as a function of space (large-$N$)}
\label{app:largeNT}
In this section, we provide the details of the derivation for Eq.~(\ref{eq:rhofM}) along with Eq.~(\ref{eq:rhof1M}).  In the large $N$ limit, 
we set {$l = N y$} and replace the discrete sum over $l$ by an integral over $y$ in Eq.~(\ref{rho_kM}). This scaling is necessary to assure that $\rho(r, N) \sim O(1)$ for large-$N$. We will check this point for self-consistency later.  Furthermore, we scale $r = z \sqrt{N}$. With this change of variable, we want to first express the summand in Eq.~(\ref{rho_kM}) as a function of {$y$} for fixed $z$ in the limit of large $N$.  Let us start with the quantity $\lambda = \sqrt{\gamma + l^2}$. Recollecting that $\gamma = c N$ and setting {$l = N y$}, we get for large $N$
\begin{equation}
\label{lambda_S}
\lambda \approx N y + \frac{c}{2y}.
\end{equation}
%{\bea \label{lambda_S}
%\lambda \approx N y + \frac{c}{2y}.
%\eea}
%\textbf{make x and y and u as x}
Approximating the Gamma function in Eq.~(\ref{rho_kM}) by the Stirling formula {$\Gamma(w+1)\sim \sqrt{2\pi}\, e^{\big(w+\frac{1}{2}\big) \log (w)-w}$ for any large argument $w$} and setting $r = z \sqrt{N}$, we find to leading order in $N$
\begin{widetext}
{\bea \label{rho_k2}
\rho_k^{\rm bulk}(r,N) \approx \frac{\sqrt{N}\Gamma(k+1)}{\pi \sqrt{2 \pi}} \int_{-\infty}^{+\infty} \frac{dy}{\sqrt{y}} \frac{e^{N \left[ y \ln (z^2/y) + y -z^2\right]} (N y)^{-k} [L_k^{N y}(z^2 N)]^2}{1+ e^{\beta (2k+1+\frac{My}{2}+ \frac{c}{2y} -\mu_\beta )}} \;.
\eea}
\end{widetext}
%Henceforth, we will absorb the $1$ in the exponent of the denominator in Eq.~\eqref{rho_k2} to redefine $\mu_\beta$ for the sake of convenience.  
We will now explain how we obtain the denominator in the integrand of above Eq.~(\ref{rho_k2}). %This is achieved by introducing an $O(1)$ parameter $M$ such that 
Note that, in the limit $\nu \to 1$, we get [using Eq.~\eqref{def_cM}]
\begin{equation}
1-\nu \sim \frac{M}{2N},\quad \text{when}\,\,\,\, \nu\to 1.
\label{eq:MNlim}
\end{equation}
Therefore, Eq.~\ref{eq:meigenvaluesgM} can be simplified as  
\begin{eqnarray}
E_{k,l} &=& 2k+1+\sqrt{\gamma+l^2}- \nu l  \nonumber \\ &=&
2k+1+\sqrt{c N+N^2 y^2}- \nu Ny \nonumber \\ &\approx&
2k+1+\frac{My}{2} +\frac{c}{2y}.
\label{eq:meigenvaluesgsim}
\end{eqnarray}
Eq.~(\ref{eq:meigenvaluesgsim}) yields the denominator in integrand of Eq.~(\ref{rho_k2}). Note that the chemical potential $\mu_\beta$ is still dependent on temperature. In the large $N$ limit, the {integral over $y$ in Eq~(\ref{rho_k2})  is dominated by a saddle point at $y=z^2$.} Therefore, it is natural to make the change of variable (from $y$ to $x$)
{\bea \label{change}
y=z^2 + \sqrt{\frac{2}{N}} \, x\, z \;.
\eea}
Hence, we get [multiplying Eq.~\eqref{change} by $N$ on both sides and simplifying]
\begin{equation}
\label{eq:nz2}
Nz^2 \approx N y - x \sqrt{2 N y}. 
\end{equation}
The choice of coefficients in Eq.~(\ref{change}) or  Eq.~(\ref{eq:nz2}) was made so that we can use the below remarkable limiting formula for the generalized Laguerre polynomials
%We can now use the following remarkable limiting formula for the generalized Laguerre polynomials,
{\bea \label{hermite}
\lim_{\lambda \to \infty} \lambda^{-k/2} L_k^\lambda(\lambda - \sqrt{2 \lambda} x) = \frac{2^{-k/2}}{\Gamma(k+1)} H_k(x) \;,
\eea}
where {$H_k(x)$} is the Hermite polynomial of index $k$. {Substituting $\lambda \approx N y$ [see Eq.~(\ref{lambda_S})] and using Eq.~\eqref{eq:nz2}
%$Nz^2 \approx N y - x \sqrt{2 N y}$ 
we find, using Eq.~\eqref{hermite}, that 
\bea \label{limit_forme_S}
\lim_{N \to \infty} (N\,y)^{-k} [L_k^{N y}(N y - x \sqrt{2 N y})]^2   = \frac{2^{-k}}{[\Gamma(k+1)]^2} H_k^2(x) \;.\nonumber \\
\eea }
Thus, the integral in Eq.~(\ref{rho_k2}) reads
\begin{eqnarray} \label{rho_k3}
\rho_k^{\rm bulk}(r,N) &\approx&  \frac{2^{-k}}{\pi^{3/2} \Gamma(k+1) } \times \nonumber \\  && \int_{-\infty}^{+\infty} dx \, \frac{e^{-x^2} \, [H_k(x)]^2}{{1+ e^{\beta (2k+1+\frac{Mz^2}{2}+ \frac{c}{2z^2} -\mu_\beta )}}} \;.\nonumber \\ %\;, \; {\rm where} \;\; a_{\pm}(k) = \frac{(\lambda_\pm(k)-z^2)\sqrt{N}}{z \sqrt{2}} \;, 
\end{eqnarray}
Note that the denominator in the integrand of Eq.~(\ref{rho_k3}) comes from the following relation [using Eq.~(\ref{change})]
\begin{equation}
\frac{My}{2} +\frac{c}{2y} \approx \frac{Mz^2}{2} +\frac{c}{2z^2}. 
\end{equation}

We recollect the normalization condition satisfied by Hermite polynomials to be 
\begin{equation}
 \int_{-\infty}^{\infty} dx \, e^{-x^2} \, [H_k(x)]^2 = 2^k  \Gamma(k+1) \sqrt{\pi}.
 \label{eq:h_norm}
\end{equation}

%\textbf{Hermite Normalization}
Using Eq.~(\ref{eq:h_norm}), we can write the final form of Eq.~(\ref{rho_k3}) as follows,
\begin{eqnarray}\label{eq:rho_k}
\rho_k^{\rm bulk}(r, N) = f_k^{\rm bulk} \bigg(\frac{r}{\sqrt{N}} \bigg),
\end{eqnarray}
where
\begin{eqnarray}
\label{eq:fkz}
f_k^{\rm bulk}(z) = \frac{1}{\pi} \frac{1}{1+ e^{\beta (2k+1+\frac{Mz^2}{2}+ \frac{c}{2z^2} -\mu_\beta )}}.
\end{eqnarray}

Therefore, if we sum over all $k$ bands, we get
\begin{equation}
\rho^{\rm bulk}(r, N) =f^{\rm bulk} \bigg(\frac{r}{\sqrt{N}} \bigg),
\label{eq:rhof}
\end{equation}
where
\begin{eqnarray}
f^{\rm bulk}(z) &=& \sum_{k=0}^{\infty}f^{\rm bulk}_k(z)  \nonumber \\
&=& \frac{1}{\pi} \sum_{k=0}^{\infty}  \frac{1}{1+ e^{\beta (2k+1+\frac{Mz^2}{2}+ \frac{c}{2z^2} -\mu_\beta )}}.
\label{eq:rhof1}
\end{eqnarray}

Note that $\mu_\beta$ in Eq.~\eqref{eq:rhof1} is fixed by the normalization condition 
\begin{eqnarray}
2\pi \int_{0}^{\infty} dz\, z\, f^{\rm bulk}(z)=1.
\label{eq:mu_norm}
\end{eqnarray}

We will end this section by discussing both the small and the high temperature limits of this expression [Eq.~(\ref{eq:rhof1})]. 

\subsection{Zero temperature limit}

In this limit ($\beta \to \infty$), for a non-zero density contribution, the term in the bracket of Eq.~(\ref{eq:fkz}) should be negative (defining $\lim_{\beta\to\infty} \mu_\beta \equiv \mu_\infty$)
\begin{equation}
\label{eq:zeq:lambda}
2k+1+\frac{Mz^2}{2}+ \frac{c}{2z^2} -\mu_\infty < 0, \,\, \text{for each }\, k 
\end{equation}
This immediately yields
\begin{eqnarray}
\label{eq:zeq:lambda1}
\sqrt{\lambda_{-} (k)} < z < \sqrt{\lambda_{+} (k)},
\end{eqnarray}
where
\begin{eqnarray} \label{lpmk}
\lambda_{\pm}(k) &=& \frac{(\mu_\infty - 2 k-1) \pm \sqrt{(\mu_\infty-2k-1)^2 - c M}}{M}.\nonumber \\ 
%k^* &=& {\rm Int}\left[ \frac{\mu - \sqrt{cM}}{2}\right] \;.
\end{eqnarray}
This reproduces the zero temperature results of Ref.~\cite{kulkarni2021multilayered} where the density in each band was given by indicator functions. In other words, in the $\beta \to \infty$ limit, Eq.~\eqref{eq:rho_k} becomes
\begin{eqnarray}
\label{eq:zeroTdensity}
\rho^{\rm bulk}_k (r,N) = \frac{1}{\pi} \mathcal{I}_{\sqrt{{\lambda_-(k)}}<z<\sqrt{{\lambda_+(k)}}}\;,
\end{eqnarray}
where the indicator function $\mathcal{I}$ takes the value $1$ if the inequality in the subscript of Eq.~\eqref{eq:zeroTdensity} is satisfied and $0$ otherwise.  
\subsection{High temperature limit}

We now discuss the high temperature limit. We start with Eq.~(\ref{eq:rhof}) and Eq.~(\ref{eq:rhof1}). Note that in the high temperature limit ($\beta \to 0$), the summation in  Eq.~(\ref{eq:rhof1}) can be replaced by an integral. Let us introduce $\beta k  = q$. Then Eq.~(\ref{eq:rhof}) becomes

\begin{eqnarray}
\rho^{\rm bulk}(r, N) &=&  \frac{1}{\pi \beta} \int_0^{\infty} dq  \frac{1}{1+ e^{\beta (\frac{Mz^2}{2}+ \frac{c}{2z^2} -\mu_\beta ) +2 q}} \nonumber \\
&=& \frac{1}{2\pi\beta} \log \big[  1 + \Delta_\beta e^{-\beta \big( \frac{Mz^2}{2} +  \frac{c}{2z^2}  \big)}\big].\nonumber\\
\label{eq:rholargeTA}
\end{eqnarray}
where $\Delta_\beta = e^{\beta \mu_\beta}$. In appendix~\ref{app:highTmu}, we argue in detail that $\Delta_\beta$ is small (in the large temperature limit) facilitating the expansion of the logarithm in Eq.~(\ref{eq:rholargeTA}), i.e., $\log[1+\epsilon] \approx \epsilon$ for some small $\epsilon$. 

In the high temperature limit, the density then becomes
\begin{equation}
\rho(r, N)  = \frac{\Delta_\beta}{2\pi\beta} e^{-\beta \big( \frac{Mz^2}{2} +  \frac{c}{2z^2}  \big)}, \text{ for}\,\, \beta \to 0. 
\label{eq:highTclassical}
\end{equation}
%with $A(\beta, M, c)$ determined by Eq.~(\ref{eq:AbetaMC}). 
Note that $\mu_\beta$ in $\Delta_\beta$ is obtained using the normalization condition Eq.~\eqref{norm_rhoN}. We therefore successfully recover the expectation of classical Gibbs-Boltzmann distribution for external potential in the high temperature limit [Eq.~(\ref{eq:highTclassical})].\\

\begingroup
\setlength{\tabcolsep}{6pt} % Default value: 6pt
\renewcommand{\arraystretch}{1.8} % Default value: 1
\begin{table}[H]
    \centering
    \begin{tabular}{ p{4cm}p{4cm}  }
        \hline
        \multicolumn{2}{c}{Some important results, parameters and their orders } \\
        \hline \hline
           $\nu =\Omega/\omega$   &   $\nu \sim 1$ \\
              $\gamma =c N$   & $c \sim O(1)$\\
        $M = (1-\nu^2) N$   & $M \sim O(1)$\\
        $r=z\sqrt{N}$ & $z\sim O(1) $\\
        $\mu_\beta \sim O(1) $& From Eq.~(\ref{eq:mu_norm}) and Eq.~(\ref{eq:rhof1})\\
        $\lambda_{\pm} (k) \sim O(1) $& From Eq.~(\ref{lpmk})\\
        $u \sim O(1)$& Edge variable [see Eq.~(\ref{eq:zuedge}), Eq.~(\ref{eq:zuedger01})]\\
         Large-$N$ density & Eq.~(\ref{eq:rhof}) and Eq.~(\ref{eq:rhof1}) \\
        Edge density & Eq.~(\ref{rho_edge_S_FTO}) and Eq.~(\ref{eq:fedge_FTO})  \\
        Normalization condition & Eq.~\eqref{eq:mu_norm}\\
           %$\gamma_\phi$& 20 - 200kHz\\
        
        \hline
        
    \end{tabular}  
    \caption{Summary of some results and typical order of parameters.}    
    \label{tab:parameters}
\end{table}
\endgroup

\section{Edge density as a function of space (large-$N$)}
\label{app:edge}

In this section, we will investigate the density behaviour at the ``edges" of the smeared wedding cake. To do so, we will start with Eq.~(\ref{rho_k2}). We will make the following change of the integrating variable ($y$) in Eq.~\eqref{rho_k2} from $y$ to $x$
{\bea \label{change_e}
y=z^2 + \sqrt{\frac{2}{N}} \, x\, z \;.
\eea}
Eq.~(\ref{rho_k2}) then becomes [using Eq.~(\ref{change_e})]
\begin{widetext}
\begin{eqnarray} \label{rho_k3_e}
\rho_k^{\rm edge}(r,N) &\approx&  \frac{2^{-k}}{\pi^{3/2} \Gamma(k+1) }  \int_{-\infty}^{+\infty} dx \, \frac{e^{-x^2} \, [H_k(x)]^2}{{1+ e^{\beta \big[2k+1+\frac{M}{2}\big(z^2 + \sqrt{\frac{2}{N}}xz\big)+ \frac{c}{2 \big(z^2 + \sqrt{\frac{2}{N}}xz\big)} -\mu_\beta \big]}}} \;.  %\;, \; {\rm where} \;\; a_{\pm}(k) = \frac{(\lambda_\pm(k)-z^2)\sqrt{N}}{z \sqrt{2}} \;, 
% &\approx& \frac{2^{-k}}{\pi^{3/2} \Gamma(k+1) }  \int_{-\infty}^{+\infty} dx \, \frac{e^{-x^2} \, [H_k(x)]^2}{{1+ e^{\beta \big[2k+1+\frac{M}{2}\big(z^2 + \sqrt{\frac{2}{N}}xz\big)+ \frac{c}{2 \big(z^2 + \sqrt{\frac{2}{N}}xz\big)} -\mu_\beta \big]}}} 
\end{eqnarray}
\end{widetext}
\subsection{Left/inner edge}
\label{subsec:left}
We will first discuss the left/inner edge.  Since we are interested in the edge, we go to new variable $u\sim O(1)$ in the following manner, 
\begin{equation}
z^2 = \lambda_-(k) + \sqrt{\frac{2  \lambda_-(k)}{N}} u\;,
\label{eq:zuedge}
\end{equation} 
where $\lambda_-(k)$ is given in Eq.~(\ref{lpmk}). We can of course, follow the same procedure for the right edge also [Appendix.~\ref{subsec:right}]. Let us elaborate Eq.~(\ref{eq:zuedge}) in terms of the original variable $r$. Eq.~(\ref{eq:zuedge}) implies, 
\begin{equation}
r^2 = \lambda_-(k) N + \sqrt{2  \lambda_-(k) N} u\;,
\label{eq:zuedger0}
\end{equation}
which essentially means 
\begin{equation}
r \approx  \sqrt{\lambda_-(k) N} + \frac{u}{\sqrt{2}},
\label{eq:zuedger01}
\end{equation}
where $u\sim O(1)$. This means that probing the edge implies going to a distance of $\sqrt{\lambda_-(k) N}$ and then zooming in at $u\sim O(1)$. From Eq.~(\ref{eq:zuedge}) one can show that [this is needed to simplify Eq.~(\ref{rho_k3_e}) further]
\begin{equation}
z^2 + \sqrt{\frac{2}{N}}xz = \lambda_-(k) + \sqrt{\frac{2  \lambda_-(k) }{N}} (u+x)\;,
\label{eq:zxu}
\end{equation}
and 
\begin{equation}
\frac{1}{z^2 + \sqrt{\frac{2}{N}}xz }= \frac{1}{\lambda_-(k)}\bigg( 1 - \sqrt{\frac{2}{\lambda_-(k) N}} (u+x) \bigg) \;.
\label{eq:zxu_inv}
\end{equation}
Plugging in Eq.~(\ref{eq:zxu}) and Eq.~(\ref{eq:zxu_inv}) in Eq.~(\ref{rho_k3_e}), we get
\bea \label{rho_edge_S_FTO}
\rho_k^{\rm inner-edge}(r,N) \to  f^{\rm inner-edge}_k(u,N)\;,%\;, \; {\rm where} \;\; f_k^{\rm edge}(u) = \frac{2^{-k}}{\pi^{3/2} \Gamma(k+1)} \int_{-u}^{\infty} dx \, e^{-x^2} \, [H_k(x)]^2 \;}
\eea
where
\begin{widetext}
\begin{eqnarray} \label{rho_k3_e_simp}
f^{\rm inner-edge}_k(u,N) %&\approx&  \frac{2^{-k}}{\pi^{3/2} \Gamma(k+1) }  \int_{-\infty}^{+\infty} dx \, \frac{e^{-x^2} \, [H_k(x)]^2}{{1+ e^{\beta \big[2k+1+\frac{M}{2}\big(z^2 + \sqrt{\frac{2}{N}}xz\big)+ \frac{c}{2 \big(z^2 + \sqrt{\frac{2}{N}}xz\big)} -\mu_\beta \big]}}} \nonumber \\ %\;, \; {\rm where} \;\; a_{\pm}(k) = \frac{(\lambda_\pm(k)-z^2)\sqrt{N}}{z \sqrt{2}} \;, 
&\approx& \frac{2^{-k}}{\pi^{3/2} \Gamma(k+1) }  \int_{-\infty}^{+\infty} dx \, \frac{e^{-x^2} \, [H_k(x)]^2}{{1+ e^{\beta \big(2 k +1 +\frac{M}{2}\lambda_-(k)+\frac{c}{2\lambda_-(k)} -\mu_{\beta}\big)} e^{\big( \frac{\beta}{\sqrt{2N}}(u+x) \sqrt{\lambda_-(k)} \big[M - \frac{c}{\lambda_-(k)^2} \big] \big)}}}\;. 
\end{eqnarray}
\end{widetext}

\textit{Zero temperature limit for the left/inner edge:}
In Eq.~(\ref{rho_k3_e_simp}), note that the denominator has two exponential pieces. Let us compute the zero temperature limit to see if we recover previous results in  Ref.~\cite{kulkarni2021multilayered}. In the limit $\beta \to \infty$, the first exponent becomes $1$ because 
\begin{eqnarray}
\lim_{\beta\to\infty} \big[ 2 k +1 +\frac{M}{2}\lambda_-(k)+\frac{c}{2\lambda_-(k)} -\mu_{\beta}\big] = 0.\nonumber \\ 
\label{eq:lmeq}
\end{eqnarray}
In fact, Eq.~(\ref{eq:lmeq}) is exactly how one of the roots of $\lambda(k)$, i.e, $\lambda_-(k)$ is determined [see also Eq.~\eqref{eq:zeq:lambda} and Eq.~\eqref{eq:zeq:lambda1}]. 
Also, in $T\to 0 $ limit, the quantity in Eq.~(\ref{rho_k3_e_simp}) survives only when $x>-u$. This is because the below identity is always satisfied 
\begin{eqnarray}
M-\frac{c}{\lambda_-(k)^2} < 0. 
\label{eq:mlambdaineq}
\end{eqnarray}

Let us now prove the inequality in Eq.~\ref{eq:mlambdaineq} which can also be written as  $ \lambda_-(k)^2 <\frac{c}{M}$. This further implies [using Eq.~(\ref{eq:lmeq}) and setting $\tilde{\mu} = \mu_\infty-2k-1$]
\begin{eqnarray}
2\tilde{\mu}^2 - Mc - 2 \tilde{\mu} \sqrt{\tilde{\mu}^2 - Mc} &<& Mc \nonumber \\
\implies 2 (\tilde{\mu}^2 - Mc) - 2 \tilde{\mu} \sqrt{\tilde{\mu}^2 - Mc} &<& 0\nonumber \\
\implies 2 \sqrt{\tilde{\mu}^2 - Mc}  \big[\sqrt{\tilde{\mu}^2 - Mc} - \tilde{\mu}\big] &<& 0
\label{eq:set_ineq}
\end{eqnarray}
The last line in Eq.~(\ref{eq:set_ineq}) is always true and hence we have proved the inequality in Eq.~(\ref{eq:mlambdaineq}). Due to this inequality, the zero temperature limit of Eq.~(\ref{rho_k3_e_simp}) finally becomes 
\bea \label{rho_edge_S_zeroT}
\rho_k^{\rm inner-edge}(r,N) \to  f^{\rm inner-edge}_k(u), \,\, \text{for }  T\to 0 \;, \nonumber \\ %\;, \; {\rm where} \;\; f_k^{\rm edge}(u) = \frac{2^{-k}}{\pi^{3/2} \Gamma(k+1)} \int_{-u}^{\infty} dx \, e^{-x^2} \, [H_k(x)]^2 \;}
\eea
where
\bea \label{rho_edge_S_e_zeroT}
f_k^{\rm inner-edge}(u) = \frac{2^{-k}}{\pi^{3/2} \Gamma(k+1)} \int_{-u}^{\infty} dx \, e^{-x^2} \, [H_k(x)]^2 \;.\nonumber \\ \,\,
\eea
Recall from Eq.~(\ref{eq:zuedge}) that 
\bea \label{exp_t_S_zeroT}
u = \sqrt{\frac{N}{2 \lambda_-(k)}}\left(\frac{r^2}{N} - \lambda_-(k)\right) \;.
\eea
Eq.~(\ref{rho_edge_S_e_zeroT}) is exactly the zero temperature result obtained in Ref.~\cite{kulkarni2021multilayered}. 
\subsection{Right/outer edge}
\label{subsec:right}

In this section, we discuss the right/outer edge. The procedure is similar to that followed in Appendix.~\ref{subsec:left}. The difference is that we go to new variable $u$ in the following manner, 
\begin{equation}
z^2 = \lambda_+(k) + \sqrt{\frac{2  \lambda_+(k)}{N}} u\;,
\label{eq:zuedgeO}
\end{equation} 
which essentially in the original variable $r=z\sqrt{N}$ means, 
\begin{equation}
r \approx  \sqrt{\lambda_+(k) N} + \frac{u}{\sqrt{2}},
\label{eq:zuedger01O}
\end{equation}
The final answer can be summarized as 
\bea \label{rho_edge_S_FTO}
\rho_k^{\rm outer-edge}(r,N) \to  f^{\rm outer-edge}_k(u,N)\;,%\;, \; {\rm where} \;\; f_k^{\rm edge}(u) = \frac{2^{-k}}{\pi^{3/2} \Gamma(k+1)} \int_{-u}^{\infty} dx \, e^{-x^2} \, [H_k(x)]^2 \;}
\eea
where 
\begin{widetext}
\begin{eqnarray} \label{eq:fedge_FTO}
f^{\rm outer-edge}_k(u,N) = \frac{2^{-k}}{\pi^{3/2} \Gamma(k+1) }  \int_{-\infty}^{+\infty} dx \, \frac{e^{-x^2} \, [H_k(x)]^2}{{1+ e^{\beta \big(2 k +1 +\frac{M}{2}\lambda_+(k)+\frac{c}{2\lambda_+(k)} -\mu_{\beta}\big)} e^{\big( \frac{\beta}{\sqrt{2N}}(u+x) \sqrt{\lambda_+(k)} \big[M - \frac{c}{\lambda_+(k)^2} \big] \big)}}}\;.\nonumber \\
\end{eqnarray}
%\label{eq:fedge_FT}
\end{widetext}

\textit{Zero temperature limit for the right/outer edge:} Here, we will recover the zero temperature limit of the outer edge.  In Eq.~(\ref{eq:fedge_FTO}), note that the denominator again has two exponential pieces. Let us compute the zero temperature limit to see if we recover previous results in  Ref.~\cite{kulkarni2021multilayered}. In the limit $\beta \to \infty$, the first exponent becomes $1$ because 
\begin{eqnarray}
\lim_{\beta\to\infty} \big[ 2 k +1 +\frac{M}{2}\lambda_+(k)+\frac{c}{2\lambda_+(k)} -\mu_{\beta}\big] = 0. \nonumber\\
\label{eq:lmeqO}
\end{eqnarray}
In fact, Eq.~(\ref{eq:lmeqO}) is exactly how the other root of $\lambda(k)$, i.e, $\lambda_+(k)$ is determined [see also Eq.~\eqref{eq:zeq:lambda} and Eq.~\eqref{eq:zeq:lambda1}]. 
Also, in $T\to 0 $ limit, the quantity in Eq.~(\ref{eq:fedge_FTO}) survives only when $x<-u$. This is because the below identity is always satisfied 
\begin{eqnarray}
M-\frac{c}{\lambda_+(k)^2} > 0. 
\label{eq:mlambdaineqO}
\end{eqnarray}

Let us now prove the inequality in Eq.~\ref{eq:mlambdaineq} which can also be written as  $ \lambda_+(k)^2 >\frac{c}{M}$. This further implies [using Eq.~(\ref{eq:lmeqO}) and setting $\tilde{\mu} = \mu_\infty-2k-1$]
\begin{eqnarray}
2\tilde{\mu}^2 - Mc + 2 \tilde{\mu} \sqrt{\tilde{\mu}^2 - Mc} &>& Mc \nonumber \\
\implies 2 (\tilde{\mu}^2 - Mc) + 2 \tilde{\mu} \sqrt{\tilde{\mu}^2 - Mc} &>& 0\nonumber \\
\implies 2 \sqrt{\tilde{\mu}^2 - Mc}  \big[\sqrt{\tilde{\mu}^2 - Mc} + \tilde{\mu} \big] &>& 0
\label{eq:set_ineqO}
\end{eqnarray}
The last line in Eq.~(\ref{eq:set_ineqO}) is always true and hence we have proved the inequality in Eq.~(\ref{eq:mlambdaineqO}). Due to this inequality, the zero temperature limit of Eq.~(\ref{eq:fedge_FTO}) finally becomes 

\bea \label{rho_edge_S_zeroTO}
\rho_k^{\rm outer-edge}(r,\theta,N) \to  f^{\rm outer-edge}_k(u), \,\, \text{for }  T\to 0\;, \nonumber\\ %\;, \; {\rm where} \;\; f_k^{\rm edge}(u) = \frac{2^{-k}}{\pi^{3/2} \Gamma(k+1)} \int_{-u}^{\infty} dx \, e^{-x^2} \, [H_k(x)]^2 \;}
\eea
where
\bea \label{rho_edge_S_e_zeroTO}
f_k^{\rm outer-edge}(u) = \frac{2^{-k}}{\pi^{3/2} \Gamma(k+1)} \int_{-\infty}^{-u} dx \, e^{-x^2} \, [H_k(x)]^2 \;.\nonumber \\ \,\,
\eea
Recall from Eq.~(\ref{eq:zuedge}) that 
\bea \label{exp_t_S_zeroTO}
u = \sqrt{\frac{N}{2 \lambda_-(k)}}\left(\frac{r^2}{N} - \lambda_+(k)\right) \;.
\eea
Eq.~(\ref{rho_edge_S_e_zeroTO}) is exactly the zero temperature results obtained in Ref.~\cite{kulkarni2021multilayered} (although it was not explicitly written there for the right/outer-edge). 
%%%%% above to be edited

%%%% IMPORTANT DISCUSSION TO BE USED LATER
\section{Temperature dependence of the chemical potential $\mu_\beta$}
\label{app:subsecR}

\subsection{The low temperature limit $\beta \to \infty$}
\label{app:lowTmuBeta}
In this section, we discuss the behavior of the chemical potential $\mu_\beta$ given by the normalization condition [Eq.~\eqref{eq:mu_norm}] in the limit $\beta \to \infty$. 

In Eq.~\eqref{eq:fedge_FTO} it is important to make some comments regarding the first exponent in the denominator. Note that, for non-trivial spatial edge behaviour, the second exponent in the denominator of Eq.~(\ref{eq:fedge_FTO}) should survive. This means $\beta \sim O(\sqrt{N})$. If this is the case, let us see what happens to the first exponent in the denominator. For this discussion, the repulsive $\gamma$ term is of no consequence and hence we set it to $0$ (i.e., $c\to0$). Note that working with right/outer edge is convenient. We want to argue that if $\beta \sim O(\sqrt{N})$, then 
\begin{equation}
\mu_\beta = \mu_{\infty} + O(e^{-\beta})
\label{eq:muinf}
\end{equation}

If Eq.~\eqref{eq:muinf} holds, then the first exponent in the denominator of Eq.~\eqref{eq:fedge_FTO} becomes $1$ because the terms in the parenthesis is zero since 
\begin{equation}
2 k +1 +\frac{M}{2}\lambda_+(k)+\frac{c}{2\lambda_+(k)} -\mu_{\infty} = 0\;.
\end{equation}

We now need to argue that Eq.~\eqref{eq:muinf} holds. The statement essentially means that if $\beta$ is tuned from $\infty$ (i.e., if temperature is raised from $0$), the change in chemical potential is rather insignificant, i.e., it has exponential corrections $\sim e^{-1/T}$. 

Note that the normalization condition Eq.~\eqref{eq:mu_norm} can be recasted as (assuming $c\to 0$ without loss of generality)

\begin{eqnarray}
%f(z) &=& \sum_{k=0}^{\infty}f_k(z)  \nonumber \\
1&=&2\pi \frac{1}{\pi} \sum_{k=0}^{\infty} \int_0^{\infty} dz\,z\,\frac{1}{1+ e^{\beta (2k+1+\frac{Mz^2}{2} -\mu_\beta )}}
\label{eq:rhof1norm}
\end{eqnarray}
Using change of variables $p=z^2$, we get 
\begin{eqnarray}
%f(z) &=& \sum_{k=0}^{\infty}f_k(z)  \nonumber \\
1&=&\sum_{k=0}^{\infty} \int_0^{\infty} dp\frac{1}{1+ e^{\beta (2k+1+\frac{M p}{2} -\mu_\beta )}}\nonumber \\
&=& \frac{2}{M\beta}  \sum_{k=0}^{\infty} \log \big[1 + e^{-\beta(2k+1-\mu_\beta)} \big]
\label{eq:rhof1norm1}
\end{eqnarray}
We will write Eq.~\eqref{eq:rhof1norm1} in a more convenient form as
\begin{eqnarray}
%f(z) &=& \sum_{k=0}^{\infty}f_k(z)  \nonumber \\
\frac{M \beta}{2} &=& \sum_{k=0}^{\rm Int[ \frac{\mu_\beta-1}{2} ]} \log \big[1 + e^{-\beta(2k+1-\mu_\beta)} \big] \nonumber \\
&+& \sum_{k=\rm Int[ \frac{\mu_\beta-1}{2} +1]}^{\infty} \log \big[1 + e^{-\beta(2k+1-\mu_\beta)} \big]
\label{eq:rhof1norm2}
\end{eqnarray}
where ``$\rm Int$" denotes the floor function. This would further imply 
\begin{eqnarray}
%f(z) &=& \sum_{k=0}^{\infty}f_k(z)  \nonumber \\
\frac{M \beta}{2} &=& \sum_{k=0}^{\rm Int[ \frac{\mu_\beta-1}{2} ]} \log \bigg[(e^{\beta(\mu_\beta-2 k-1)})(1 + e^{-\beta(\mu_\beta-2 k-1)}) \bigg] \nonumber \\
&+& \sum_{k=\rm Int[ \frac{\mu_\beta-1}{2}+1 ]}^{\infty} \log \big[1 + e^{-\beta(2k+1-\mu_\beta)} \big]
\label{eq:rhof1norm3}
\end{eqnarray}
After some simplification (and assuming large-$\beta$ which helps us to use $\log[1+\epsilon] \approx \epsilon$ for some small $\epsilon$), we get 
\begin{eqnarray}
%f(z) &=& \sum_{k=0}^{\infty}f_k(z)  \nonumber \\
\frac{M}{2}  &=& \sum_{k=0}^{\rm Int[ \frac{\mu_\beta-1}{2} ]} (\mu_\beta - 2 k-1 )  + \frac{1}{\beta}\sum_{k=0}^{\rm Int[ \frac{\mu_\beta-1}{2} ]} e^{-\beta(\mu_\beta-2 k-1)}  \nonumber \\
&+& \frac{1}{\beta}  \sum_{k=\rm Int[ \frac{\mu_\beta-1}{2}+1 ]}^{\infty} e^{-\beta(2k+1-\mu_\beta)} 
\label{eq:rhof1norm4}
\end{eqnarray}

It is to be noted that henceforth we will assume that $\frac{\mu_\beta-1}{2}$ is not an integer.  

Let us now make the expansion (about zero temperature), 
\begin{equation}
\mu_\beta = \mu_\infty + \delta\mu
\label{eq:mu_exp}
\end{equation}

Note that at zero temperature ($\beta\to \infty$), we have~\cite{kulkarni2021multilayered}, 
\begin{equation}
\frac{M}{2} = \sum_{k=0}^{\rm Int[ \frac{\mu_\infty-1}{2} ]} (\mu_\infty - 2 k -1)  
\label{eq:zeroTmu}
\end{equation}

Using Eq.~\eqref{eq:mu_exp} and  Eq.~\eqref{eq:zeroTmu} in Eq.~\eqref{eq:rhof1norm4}, we get

\begin{equation}
\rm Int \big[ \frac{\mu_\infty-1}{2} \big] \delta \mu  + \frac{1}{\beta} e^{- A \beta }\approx 0
\label{eq:mu_exp_simp}
\end{equation}
where $A>0$ is given by

\bea \label{A_app}
A = \min\bigg\{\mu_\infty - 2 \,\rm Int \left[ \frac{\mu_\infty-1}{2}\right]-1, \\ 2\, \rm Int \left[ \frac{\mu_\infty-1}{2}\right] +3 - \mu_\infty \bigg\} \;. \nonumber \\
\eea
Therefore, the large $\beta$ behavior of $\mu_\beta$ reads
\bea \label{final_largeb}
\mu_\beta \approx \mu_{\infty} - \frac{1}{\rm Int \big[ \frac{\mu_\infty-1}{2} \big]} \frac{1}{\beta}\, e^{-A\, \beta} \;.
\eea

\subsection{High temperature limit $\beta \to 0$}
\label{app:highTmu}

We now discuss the high temperature limit of $\mu_\beta$ given by the normalization condition in Eqs. \eqref{eq:rhof1M} and \eqref{eq:mu_normM}. We first note that in the high temperature limit ($\beta \to 0$), the discrete sum in Eq.~\eqref{eq:rhof1M} can be replaced by an integral. Setting $\beta k  = q$, Eq.~(\ref{eq:rhof1M}) becomes
\begin{eqnarray}
f^{\rm bulk}(z) &\approx&  \frac{1}{\pi \beta} \int_0^{\infty} dq  \frac{1}{1+ e^{\beta (\frac{Mz^2}{2}+ \frac{c}{2z^2} -\mu_\beta ) +2 q}} \nonumber \\
&=& \frac{1}{2\pi\beta} \log \big[  1 + e^{-\beta \big( \frac{Mz^2}{2} +  \frac{c}{2z^2} - \mu_\beta \big)}\big].
\label{eq:rholargeT}
\end{eqnarray}
Injecting this Eq.~\eqref{eq:rholargeT} in the normalization condition in Eq.~\eqref{eq:mu_normM} yields 
\begin{equation} \label{asympt_1}
1= \frac{1}{\beta} \int_0^\infty  dz \, z\,  \log \bigg[  1 + e^{-\beta \big(\frac{Mz^2}{2} +  \frac{c}{2z^2} - \mu_\beta \big)}\bigg] \;.
\end{equation}
Performing a change of variable $u = \sqrt{\beta} z$ and keeping the leading terms of small $\beta$, one gets
\begin{eqnarray} \label{asympt_2}
\beta^2 &=& \int_0^\infty du \, u\, \log{\left( 1 + \Delta_\beta e^{- \frac{M}{2}u^2} \right)}\big(1 + o(1)\big) \nonumber \\
&=& - \frac{1}{M}{\rm Li}_2(-\Delta_\beta)\big(1 + o(1)\big)  \;,
\end{eqnarray}
where ${\rm Li}_2(v) = \sum_{k=1}^\infty v^k/k^2$ is the poly-logarithm function and $\Delta_\beta = e^{\beta \mu_\beta}$. Since $-{\rm Li}_2(-v)$ is a monotonically increasing function of $z$, for the right hand side of Eq. (\ref{asympt_2}) to be small [of order $O(\beta^2)$], one needs to have $\Delta_\beta$ also small. Using the small $v$ expansion $-{\rm Li}_2(-v) = v + O(v^2)$, we get $\Delta_\beta = M \beta^2$ which implies
\bea \label{asympt_3}
\mu_\beta = 2\frac{\log(\beta)}{\beta} \big(1 + o(1)\big) \;. 
\eea

\section{The case of general potential}
\label{app:generalP}

In this Appendix we give a detailed discussion for a more general potential $V(r)$. As in the case of specific potential in Eq.~\eqref{def_VM} discussed before, for a generic spherically symmetric potential $V(r)$, we can again decompose the Schrodinger equation into the angular and radial sectors. We label the angular eigenfunctions before by $l=0,\pm1,\pm 2 \cdots$. For each $l$ the radial part of the wavefunction satisfies the effective one dimensional Schrodinger equation
\bea
\label{eq:1DSE}
\hat H_l \chi_{k,l} (r) = E_{k,l} \chi_{k,l} (r) \;,
\eea
where $\hat H_l$ is the angular Hamiltonian in the sector with angular momentum $L_z=l$. In units $m= \hbar=1$, $\hat{H}_l$ reads 
\be \label{angular} 
\hat H_l = - \frac{1}{2} \bigg( \partial_r^2 + \frac{1}{r}\partial_r\bigg) + V_l(r), %,\quad V_\ell(r) = V(r) + \frac{\ell^2}{2 r^2} - \Omega \ell
\ee 
where
\be
\label{angularV} 
 V_l(r) = V(r) + \frac{l^2}{2 r^2} - \Omega.
\ee

This one-dimensional Schr\"odinger equation [Eq.~\eqref{eq:1DSE}], for fixed $l$, has energy levels which we label, as before, by $k=0,1,2\cdots$ (assuming that $V_l(r)$ is confining, i.e., $V_l(r) \to \infty $ as $r \to \infty$). Thus the $k$'s once again denote the Landau levels. At zero temperature, we occupy these levels up to the Fermi level $\mu$. In the specific example studied before, we were classifying the occupied levels for each filled Landau band index $k$ and for each $k$, the angular index $l$ was varying from $l_-(k)$ to $l_+(k)$. Here, instead, we will describe these occupied levels in the opposite direction, i.e., we fix angular index $l$ and for each $l$, the occupied $k$ levels runs from $0$ to $m_l$ where $m_l$ is the highest occupied level in the $l-\rm{th}$ angular sector. Using this labelling scheme, at zero temperature the total two dimensional density is given by its angular decomposition
\be \label{eq:2Dden}
2 \pi r \rho(r, N) =  \sum_{l = -\infty}^{+\infty} \rho_l(r) \quad , \quad \int_0^\infty  dr \rho_l(r)  = m_l \nonumber \\
\ee 
where 
\begin{equation}\label{eq:rho_chi}
\rho_l(r) = \sum_{k=0}^{m_l}|\chi_{k,l}(r)|^2 
\end{equation}
is the density of the one-dimensional problem 
[Eq.~\eqref{eq:1DSE}] with $m_l$ fermions. 

We now consider a class of potentials of the form 
\be 
V(r) = \frac{r^2}{2} + v\bigg(\frac{r}{\sqrt{N}}\bigg) \;,
\ee 
where $v(z)$ is a given smooth function. The specific case we considered before corresponds to $v(z) =c/2z^2$. 

As before the natural scale of $r$ is $\sqrt{N}$. Hence, we set
\begin{equation}
\label{eq:appR}
r = z \sqrt{N}\;,
\end{equation}
and, as before, we also set
\begin{equation}
\label{eq:appL}
l = N y \;.
\end{equation}
Hence, Eq.~\eqref{angularV} in these rescaled coordinates reads 
\be 
\label{eq:appVlr}
V_l(r) = N \bigg( \frac{z^2}{2} + \frac{y^2}{2 z^2} - \Omega y\bigg) + v(z) 
\ee 
The leading term proportional to $N$ in Eq.~\eqref{eq:appVlr}, when plotted as a function of $z$ for fixed $y$ has a minimum at $z=\sqrt{y}$. Thus the leading term of the potential around this minimum looks like a Harmonic oscillator. Hence, we expand $V_l(r)$ in Eq.~\eqref{eq:appVlr} around the minimum at $z = \sqrt{y}$ by setting 
\begin{eqnarray}
\label{eq:appz}
z = \sqrt{y} - \frac{x}{\sqrt{2N}}\;,
\end{eqnarray}
where $x \sim O(1)$ and denotes the scaled distance from the minimum of the potential. This  gives
\be \label{eq:yxz}
y = z^2 + \sqrt{\frac{2}{N} } x z + \frac{x^2}{2 N} \simeq z^2 + \sqrt{\frac{2}{N} } x z.
\ee 
We also note that in terms of $x$, we can rewrite the radial distance $r$ in Eq.~\eqref{eq:appR} as [using Eq.~\eqref{eq:appL} and Eq.~\eqref{eq:appz}]
\be 
\label{eq:apprx}
r =\sqrt{\ell} - \frac{x}{\sqrt{2}}\;. 
\ee 

In terms of variable $x$ the Hamiltonian in Eq.~\eqref{angular} reads 

\bea
\label{eq:appHV}
H_l &=& - \partial_x^2 + V_\ell(r)  \\
  \nonumber  V_l\bigg(r = \sqrt{l} - \frac{x}{\sqrt{2}}\bigg) &=& N (1- \Omega ) y  + v(\sqrt{y})  + x^2 \\ \nonumber &+& 
\frac{1}{\sqrt{2 y N}} \bigg(x^3 -  x \sqrt{y} v'(\sqrt{y})\bigg)  \\ &+& \frac{1}{N} \bigg( \frac{5 x^4}{8 y} + \frac{x^2 v''(\sqrt{y})}{4 } \bigg) + \dots \nonumber \\
\label{eq:apponlyV}
\eea
Using the fact that $1-\Omega = 1-\nu  = M/2N$ [see Eq.~\eqref{eq:MNlim}], the first three terms in Eq.~\eqref{eq:apponlyV} are $O(1)$ while the terms on the second and third line are respectively of $O(1/\sqrt{N})$ and $O(1/N)$. 
Hence, the Hamiltonian in Eq.~\eqref{eq:appHV}
is the Hamiltonian of a 1D harmonic oscillator (HO) in the variable $x$ (with $m=1/2$ and $m \omega^2 = 2$, 
hence $\alpha=\sqrt{m \omega}=1$) perturbed by a $O(1/\sqrt{N})$ cubic and $O(1/N)$ quartic term.

At leading order, one finds that the eigenenergies of $H_l$ are
given by
\be  \label{eps} 
\epsilon_{k,l} \approx \epsilon_k(y) = N (1- \Omega ) y + v(\sqrt{y})  + 2 k + 1 
\ee 
where $k =0, 1, 2\cdots $ is an integer, and we recall that $y=l/N$. Using first order perturbation theory, and 
since for the HO the matrix elements of the form $\langle n | x^{2 p+1} |n \rangle =0$ by symmetry (with $p$ being an integer), one can
show that the leading order corrections to the result in Eq. (\ref{eps}) are of order $O(1/N)$. 
\\

Using standard results about the 1D HO, and the above angular decomposition,
we can then write the density at any temperature, and to leading order in $N$
as 
\bea
&& 2 \pi r \rho(r,N) =  \sum_{l=-\infty}^{+\infty} \rho_l(r) \\
&&
\rho_\ell(r) dr \simeq \sum_{k = 0}^{\infty} 
\frac{ |\psi_k(x)|^2 }{1 + e^{\beta (\epsilon_k(y) - \mu_\beta)}} \; dx
\\
&& \psi_k(x)=\frac{1}{\sqrt{2^k k! \sqrt{\pi}}} e^{- \frac{x^2}{2}} H_k(x) \;.
\label{eq:phikherm}
\eea 
Using $|dx/dr| = \sqrt{2}$ from Eq.~\eqref{eq:apprx} and replacing the sum over $l$ by an integral over $y$ using $l=Ny$, we get  
\be 
\label{eq:apprhopsi}
2 \pi r \rho(r,N) \simeq \sqrt{2} N \int_{-\infty}^{+\infty} dy \sum_{k= 0}^{\infty} 
\frac{ |\psi_k(x)|^2 }{1 + e^{\beta (\epsilon_k(y) - \mu_\beta)}}  \nonumber \\
\ee 
where we recall that in these formulae $x=\sqrt{2 N} (\sqrt{y}-z)$
and $r=z\sqrt{N}$. Note that substituting Eq.~\eqref{eps} in Eq.~\eqref{eq:apprhopsi} together with the relation  $dy = \sqrt{\frac{2}{N} } z\,dx$ [using Eq.~\eqref{eq:yxz}], we get 
\begin{equation}
    \rho(r,N) = f(z,N)\;,
\end{equation}
where 
\be
\label{eq:fzn}
f(z,N) = \frac{1}{\pi} \sum_{k = 0}^{\infty}   \int_{-\infty}^{+\infty} dx  \frac{ |\psi_k(x)|^2 }{1 + e^{\beta (\frac{M}{2}  y + v(\sqrt{y})  + 2 k + 1  - \mu_\beta)}}\;. \nonumber \\
\ee
where $y$ as a function of $x$, for fixed $z$, is given by $y \simeq z^2 + \sqrt{\frac{2}{N} }\, x \,z$ as in Eq,~\eqref{eq:yxz}. Note that in Eq.~\eqref{eq:fzn}, we have kept terms up to $O(1/\sqrt{N})$, but neglected terms of $O(1/N)$. Also, note that the energy levels in Eq.~\eqref{eps} are also valid up to $O(1/\sqrt{N})$, since the $O(1/\sqrt{N})$ term in Eq.~\eqref{eq:appHV} does not change the energy level as argued before.
Below, we start from Eq.~\eqref{eq:fzn} and focus separately on the bulk and edge regimes. 
\begin{figure}[t]
\includegraphics[width =\linewidth]{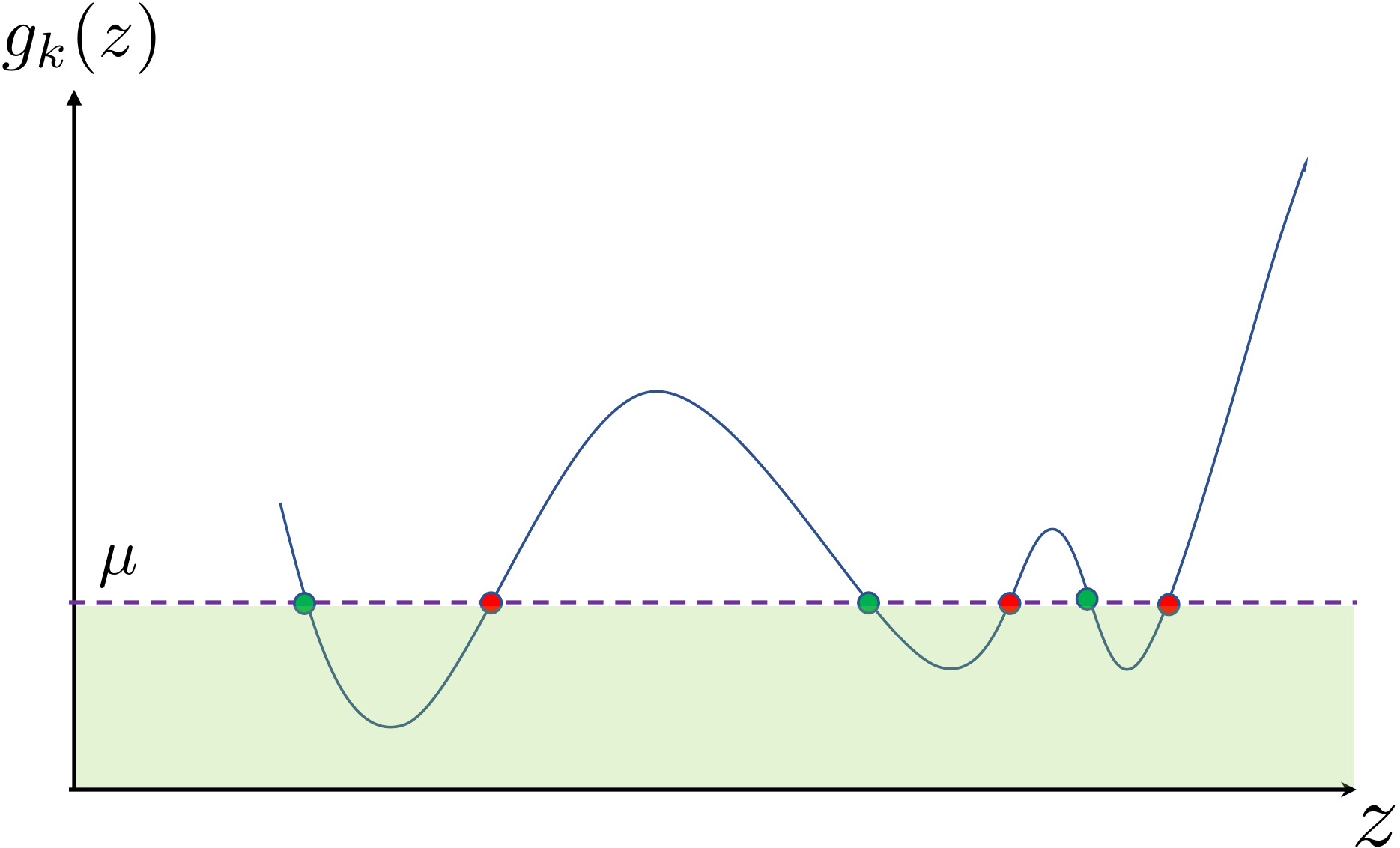}
\caption{A schematic plot of $g_k(z)$ vs. $z$ given by Eq.~\eqref{eq:gkz} for a general function $v(z)$. The support of the $k$-th layer's density
is the set of points with $g_k(z)<\mu$ (shaded green region). The filled circles denote the locations $z=z_e$ of the edges of this support where $g_k(z_e) = \mu$ (purple dashed line). It is easy to see that for every interval (which is defined as a segment between a green and a red circle), the support of the outer boundary (red) has $b>0$ [Eq.~\eqref{eq:eq:effT}] and support of the inner boundary (green) has $b<0$. }
\label{fig:app_schematic}
\end{figure}

\subsection{Bulk (general potential)}

In the bulk, we set $y=z^2$ and neglect the $O(1/\sqrt{N})$ corrections. Substituting $y=z^2$ in Eq.~\eqref{eq:fzn}, we see that the ``Fermi factor" form becomes independent of $x$ and comes out of the $x$ integral. Hence using the normalization condition $\int_{-\infty}^{+\infty} dx |\psi_k(x)|^2=1$, we find that the function $f(z,N)$ in Eq.~\eqref{eq:fzn} becomes only a function of the scaled variable $z$. Hence, we obtain
\begin{equation}
\rho^{\rm bulk}(r,N) \approx f^{\rm bulk} \bigg(\frac{r}{\sqrt{N}} \bigg),
\label{eq:rhofMA}
\end{equation}
where
\bea \label{eq:fbulkgen}
&& f^{\rm bulk}(z) = \frac{1}{\pi} \sum_{k = 0}^{\infty} 
\frac{ 1 }{1 + e^{\beta \big(g_k(z) - \mu_\beta \big)}}\;, 
\eea 
with $g_k(z)$ given by
\begin{equation}
\label{eq:gkz}
g_k(z)=\frac{M}{2} z^2 + v(z)  + 2 k + 1
\end{equation}

\noindent \textit{Zero temperature limit:} At $T=0$ one can replace the ``Fermi factor" form in Eq.~\eqref{eq:fbulkgen} by a theta function which gives Eq.~\eqref{eq:maingenT0} in the main text, namely
\bea 
\label{eq:maingenT0app}
f^{\rm bulk}(z) &=& \frac{1}{\pi}   \sum_{k = 0}^{\infty} 
\, \theta \big( \mu - g_k(z) \big) \;, % \int dx |\psi_k(x)|^2 \nonumber \\
\eea 
where $g_k(z)$ is defined in Eq.~\eqref{eq:gkz}.
%\begin{figure}[t]
%\includegraphics[width =\linewidth]{schematic_gk.pdf}
%\caption{A schematic plot of $g_k(z)$ vs. $z$ given by Eq.~\eqref{eq:gkz} for a general function $v(z)$. The support of the $k$-th layer's density
%is the set of points with $g_k(z)<\mu$ (shaded green region). The filled circles denote the locations $z=z_e$ of the edges of this support where $g_k(z_e) = \mu$ (purple dashed line). It is easy to see that for every interval (which is defined as a segment between a green and a red circle), the support of the outer boundary (red) has $b>0$ [Eq.~\eqref{eq:eq:effT}] and support of the inner boundary (green) has $b<0$. }
%\label{fig:app_schematic}
%\end{figure}
\subsection{Edge (general potential)}
We now study the large-$N$ density profile near the edges. At $T=0$ these edges $z_e=z_e(k)$ for a fixed layer $k$ are given by 
the solutions of [see Eq.~\eqref{eq:maingenT0app}]  
\begin{equation} \label{edgepositions} 
\frac{M}{2} z_e^2 + v(z_e)  + 2 k + 1  =  \mu \;, %\label{ze} 
\end{equation}
where $\mu=\mu_\infty$. Note that there can be either no solution, a single or multiple solutions to this Eq.~(\ref{edgepositions}).
We now study $f(z,N)$ in Eq.~\eqref{eq:fzn} near such edges $z_e$ where we parametrize the scaled position $z$ as
\begin{equation} 
\label{eq:zze}
z = z_e + \frac{u}{\sqrt{2 N}} \;,
\end{equation} 
where $u\sim O(1)$. 
Using Eq.~\eqref{eq:zze} the variable $y$ in Eq.~\eqref{eq:fzn} becomes for large-$N$
\bea
y &=& z^2 + \sqrt{\frac{2}{N} } x z \nonumber \\
%&& = (z_e + \frac{u}{\sqrt{2 N}} )^2 + \sqrt{\frac{2}{N} } x z 
&=& z_e^2 + \sqrt{\frac{2}{N} } (x+u)  z_e  + O\bigg(\frac{1}{N}\bigg) 
\ee 
Hence the argument inside the exponential of the ``Fermi factor" form in Eq.~\eqref{eq:fzn} becomes, using \eqref{edgepositions}
\bea \label{eq:appze}
&& \frac{M}{2}  y + v(\sqrt{y})  + 2 k + 1  - \mu_\beta  \nonumber \\ 
&& = \mu - \mu_\beta + \frac{1}{2} \big(M z_e + v'(z_e) \big) \sqrt{\frac{2}{N} } (x+u) %\nonumber \\
%&& 
+ O\bigg(\frac{1}{N}\bigg)\nonumber \\
\eea 
We emphasize, as discussed before below Eq. (\ref{eps}), that there is no $O(1/\sqrt{N})$ correction coming from $\epsilon_{k,l}$ in Eq.~(\ref{eps}). This ensures that the neglected terms in Eq. (\ref{eq:appze}) are indeed of order $O(1/N)$. Using Eq.~\eqref{eq:appze} in 
Eq.~\eqref{eq:fzn} we get for a fixed layer $k$
\bea\label{eq:app_rho_f}
\rho_k(r,N) &=& f_k\bigg(z=z_e+\frac{u}{\sqrt{2N}},N\bigg)  \nonumber \\& \approx&
\frac{1}{\pi}   \int_{-\infty}^{+\infty} dx  
\frac{ |\psi_k(x)|^2 }{1 + e^{\beta (\mu - \mu_\beta)} e^{\frac{\beta}{\sqrt{2 N}}  \big(M z_e + v'(z_e) \big)  (x+u) }} \nonumber \\
\eea 
where we recall that $z_e$ is any one of the roots of Eq.~\eqref{edgepositions}. For this expression to approach an $N$-independent form
as $N \to \infty$, it is natural to scale $\beta \sim O(\sqrt{N})$. In this scaling limit, setting $\beta = \tilde \beta \sqrt{2N}$ implies 
$\mu_\beta \to \mu_\infty \equiv \mu$ as $N \to \infty$. We then expect (assuming we are not exactly at the nucleation/formation point of a new layer) that $\beta (\mu - \mu_{\beta}) \to 0$ as $N \to \infty$ for a general $v(z)$, as shown before in Appendix.~\ref{app:lowTmuBeta} for the special case $v(z) = 0$. Therefore, the density at the edge in Eq.~\eqref{eq:app_rho_f} for a fixed layer $k$ takes the scaling form

\begin{widetext}
\begin{eqnarray} \label{eq:fedge_FTMGenapp}
f^{\rm edge}_k(u) = \frac{2^{-k}}{\pi^{3/2} \Gamma(k+1) }  \int_{-\infty}^{+\infty} dx \, \frac{e^{-x^2} \, [H_k(x)]^2}{{1+ e^{ \tilde{\beta}(u+x) \big(Mz_e+v^{\prime}(z_e)\big) }}},
\end{eqnarray}
\end{widetext}
where we have used Eq.~\eqref{eq:phikherm} and $z_e$ is given by Eq.~\eqref{edgepositions}. Note that for the specific case of $v(z)=c/2z^2$ we find $z_e = \sqrt{\lambda_{-}(k)}$ and $z_e = \sqrt{\lambda_{+}(k)}$ for the left and right edge respectively where $\lambda_{\mp}(k)$ are given in Eq.~\eqref{lpmk_txte}. Hence, 
\begin{equation}
Mz_e+v^{\prime}(z_e) =  \sqrt{\lambda_{\mp}(k)}\bigg(M-\frac{c}{\lambda_{\mp}(k)^2}\bigg)\;,
\end{equation}
which together with Eq.~\eqref{eq:fedge_FTMGenapp} yields back the expression given in the text in Eq.~\eqref{eq:fedge_FTMsimp}. Note that, as in the specific example above, one can define  an ``effective scaled inverse temperature" $b$ as
\be 
\label{eq:eq:effT}
b = \frac{\beta}{\sqrt{2 N}} g_k'(ze)= \frac{\beta}{\sqrt{2 N}}  \big( M z_e + v'(z_e) \big) = \frac{\beta}{\sqrt{2 N}}  \frac{d\mu}{dz_e}\;, \nonumber \\
\ee 
where in the last equality in Eq.~\eqref{eq:eq:effT} we have used 
\begin{equation}
    \frac{d\mu}{dz_e} = M z_e + v'(z_e)\,, 
\end{equation}
which is obtained by differentiating Eq.~\eqref{edgepositions} with respect to $z_e$. In terms of $b$ the scaling function $f_k^{\rm edge}(u)$ in Eq.~\eqref{eq:fedge_FTMGenapp} takes a universal form

\begin{eqnarray} \label{eq:fedge_FTMGenappuniv}
f^{\rm edge}_k(u) = \frac{2^{-k}}{\pi^{3/2} \Gamma(k+1) }  \int_{-\infty}^{+\infty} dx \, \frac{e^{-x^2} \, [H_k(x)]^2}{{1+ e^{ b(u+x) }}},\nonumber\\
\end{eqnarray}
which is independent of the potential $v(z)$. For the special case $v(z)=c/2z^2$, we recover  Eq.~\eqref{eq:fedge_FTMsimp} and Eq.~\eqref{eq:bpm}.

Let us recall from Eq.~\eqref{eq:maingenT0app} that inside a given layer $g_k(z)<\mu$ and outside the layer $g_k(z)>\mu$. Hence, $g_k(z=z_e) = \mu$ fixes the edges of a given layer as in Eq.~\eqref{edgepositions}. 

This support formed by the edges of a given layer $k$ can be either empty, or a single interval, or multiple intervals (we assume that $g_k(z)$ is smooth and differentiable and growing at infinity), depending on the specific form of $v(z)$, and hence that of $g_k(z)$. A schematic plot of $g_k(z)$ versus $z$ is shown in Fig.~\ref{fig:app_schematic}. 
Clearly when $\mu$ is increased each interval of the support increases in size. The edges of these intervals are the solutions of
$g_k(z_e)=\mu$ and the derivatives $g_k'(z_e)$ thus have alternating signs. Hence, from Eq.~\eqref{eq:eq:effT}, for each interval of the support
the outer boundary has $b>0$ and the inner boundary has $b<0$. \\
 
\noindent \textit{Zero temperature limit:} 
To study the $T=0$ limit, we see from Eq.~\eqref{eq:fedge_FTMGenappuniv} that the two cases $b>0$ (outer edges) and $b<0$ (inner edges) must be treated separately. Indeed, for $b>0$ in the $T\to 0$ limit we have 
\begin{eqnarray} \label{eq:fedge_bplus}
f^{\rm outer-edge}_k(u) = \frac{2^{-k}}{\pi^{3/2} \Gamma(k+1) }  \int_{-\infty}^{-u} dx \, e^{-x^2} \, [H_k(x)]^2 \;.\nonumber\\
\end{eqnarray}
This result is valid for any potential potential $v(z)$ and it 
coincides with the result obtained for the specific case of $v(z) = c/2z^2$ in Eq.~\eqref{rho_edge_S_e_zeroTO}. Similarly, for $b<0$, in the $T\to 0$ limit we have 
\begin{eqnarray} \label{eq:fedge_bminus}
f^{\rm inner-edge}_k(u) = \frac{2^{-k}}{\pi^{3/2} \Gamma(k+1) }  \int_{-u}^{+\infty} dx \, e^{-x^2} \, [H_k(x)]^2 \;,\nonumber\\
\end{eqnarray}
which also is valid for any potential and coincides with the result obtained for the specific case of $v(z) = c/2z^2$ in Eq.~\eqref{rho_edge_S_e_zeroT}. 

\bibliographystyle{apsrev4-1}
\bibliography{refs.bib}

\end{document}